\documentclass[twocolumn,revtx4,iop,apj]{emulateapj}
\usepackage{longtable}
\usepackage{aas_macros}
\usepackage{graphicx}
\usepackage{amssymb}
\usepackage{natbib}
\usepackage{url}
\newcommand{\Bl}{B_{\rm los}}

\defcitealias{trt_p2}{Paper II}
\defcitealias{trt_p3}{Paper III}

\shorttitle{Helioseismology of Pre-Emerging Active Regions I}
\shortauthors{Leka et al.}

\begin{document}

\title{Helioseismology of Pre-Emerging Active Regions I: Overview, Data, and Target Selection Criteria}
\author{K.D.~Leka\altaffilmark{1},  G.~Barnes\altaffilmark{1},  A.C.~Birch\altaffilmark{1,3}, I.~Gonzalez-Hernandez\altaffilmark{2}, T.~Dunn\altaffilmark{1},  B.~Javornik\altaffilmark{1}, D.C.~Braun\altaffilmark{1}}

\altaffiltext{1}{NorthWest Research Associates, Boulder, CO 80301 USA}
\altaffiltext{2}{National Solar Observatory, Tucson, AZ 85719 USA}
\altaffiltext{3}{Max-Planck Institut f\"{u}r Sonnensystemforschung, 37191 Katlenburg-Lindau,  Germany}

\begin{abstract}

This first paper in a series describes the design of a study testing whether pre-appearance signatures of solar magnetic active regions were detectable using various tools of local helioseismology. The ultimate goal is to understand flux- emergence mechanisms by setting observational constraints on pre-appearance subsurface changes, for comparison with results from simulation efforts. This first paper provides details of the data selection and preparation of the samples, each containing over 100 members, of two populations: regions on the Sun that produced a numbered NOAA active region, and a ÒcontrolÓ sample of areas that did not. The seismology is performed on data from the GONG network; accompanying magnetic data from SOHO/MDI are used for co-temporal analysis of the surface magnetic field. Samples are drawn from 2001 -- 2007, and each target is analyzed for 27.7 hr prior to an objectively determined time of emergence. The results of two analysis approaches are published separately: one based on averages of the seismology- and magnetic-derived signals over the samples, another based on Discriminant Analysis of these signals, for a statistical test of detectable differences between the two populations. We include here descriptions of a new potential-field calculation approach and the algorithm for matching sample distributions over multiple variables. We describe known sources of bias and the approaches used to mitigate them. We also describe unexpected bias sources uncovered during the course of the study and include a discussion of refinements that should be included in future work on this topic.

\end{abstract}


\maketitle

\section{Introduction}
\label{sec:intro}

We refer to the appearance of new solar active regions as ``emergence'',
implying a rise from below the visible photosphere.  Yet the appearance
and evolution of an active region from the surface through the corona is
the symptom, the result -- filtered through the $\tau=1$ boundary and the
transitions from high- to low-$\beta$ plasmas -- of some (yet unknown)
process happening below the visible surface.

One general class of theories suggests that active regions form as the result
of magnetic flux concentrations rising buoyantly from the base of
the convection zone \citep[for a review see][]{Fan2009}.  Another
possibility is that sunspots are formed via coagulation of magnetic
fields generated closer to the solar surface \citep[][and references
therein]{Brandenburg2005}.  The pre-emergence seismic signatures expected
from these two approaches differ substantially.   From the former 
scenario one should expect signals generally taking the form
of a bulk and quickly moving disturbance whose internal plasma flow should
result in a signal detectable with today's tools \citep{BirchBraunFan10}.  
In the latter case, the expectation would likely be a slower change in the sub-surface temperature,
flow, and magnetic field environment over a less localized area.   Simulations which focus on 
the dynamics of flux systems rising through the upper layers 
imply that slowly rising flux systems may impact the convection only 
minimally \citep{stein_etal_2011}, depending on the field strengths involved.
Still, simulations provide clues but are limited; observations must continue
to provide guidance.

Being able to peer below the visible surface at the sub-surface structure
and dynamics could provide the guidance regarding the formation mechanism for
solar active regions (``AR'').  Helioseismology seems to
promise the ability to detect changes in the flow patterns and temperature
beneath the visible surface.  From the pure physics perspective, the tools
of local helioseismology \citep{Gizon2005,Gizon2010} should help determine
the subsurface dynamics associated with active region formation, and thus
could provide evidence for or against the basic model types.
Some preliminary work (described below) applying sensitive tools of this type to data-sets
well suited for these techniques suggests that the capability may now
be available.

Most recent efforts have been case studies,
focusing on the emergence of one or a few active regions
\citep[e.g.,][]{jensen_etal_2001,zharkovthompson2008,komm_etal_2008,ilonidis_etal_2011,Braun2012}.
The results have been inconclusive when taken as an ensemble, possibly
due to the physics of active region emergence, possibly due to the 
differences between the studies themselves.
Inverting time-distance data from MDI using three-dimensional kernels,
\citet{jensen_etal_2001} found perturbations indicating wave-speed increases
20\,Mm below two active regions in the hours after their appearance.
\citet{zharkovthompson2008}, using a very similar method for two active regions, found a similar
increase when surface flux was visible, but also a ``loop-like structure''
with decreased sound-speed, days prior to the appearance of surface flux.
\citet{ilonidis_etal_2011} also employ time-distance analysis of MDI data, and
present very large negative travel-time shifts (increases in the sound speed) 
located between 42-75\,Mm up to two days prior to surface flux appearance of four active regions.  
They associate these disturbances
with magnetic structures emerging at speeds of 0.3-0.6 km s${}^{-1}$, and do see
a high rate of flux emergence following the perturbations.  Yet \citet{Braun2012}
using acoustic holography on the same data for the same four active regions, 
detect no such unique signals at the specified times and depths.
Employing ring-diagram analysis of GONG data for 13 new or growing active regions
(and contrasting with control areas), \citet{komm_etal_2008}
found evidence for upflows prior to the appearance of emerging flux 
at the surface, followed by a transition to predominantly downflows 
once the active region was established.

Ring-diagram analysis was also used in statistical studies of
seismic signatures associated with emerging magnetic flux,
comparing average signals for hundreds of regions with increasing
flux to either ``quiet'' areas or to those with decreasing flux
\citep{komm_etal_2009,komm_etal_2011}.  While the analysis had fairly low
temporal and spatial resolution, upflows were associated with emerging
flux at depths below 10Mm whereas at shallower layers, upflows changed to
downflows as surface field became stronger.  These studies examined
the broad spectrum of surface-field behavior: growing flux, consistent flux, and
decreasing flux.  However, the ``emerging flux'' category did not differentiate between
``new'' active regions and emerging flux within already
established regions.

The conflicting results in case studies could indicate that there is no
unique signature, or that results are sensitive to subtle methodology
differences.  The few published statistical studies have been based
on a single method, and now need to be refined to focus solely on the
pre--emergence context, and employ higher resolution analysis.

In the present investigation we employ a combination of local
helioseismology, surface magnetic field diagnostics and statistical tests
to examine 
what can be learned with regards to sub-surface magnetic flux systems,
their structure, and their evolution.
The basic premise of this series of papers \citep[this paper along
with][]{trt_p2,trt_p3} is to determine if there are detectable changes
in the solar interior that indicate an emerging active region prior to
the appearance at the solar surface of a magnetic field concentration.

We have designed and completed a study to examine the possibility
of pre--emergence detection of active regions, with the goal of
characterizing the sub-surface changes in the context of emerging-flux
models.  The approach pays attention to sources of bias, statistical
and systematic error, and includes statistical validation of the
results.  The organization of this paper is as follows: in section
\S~\ref{sec:design} we outline the physical parameters within which the
overall study must work, and the statistical motivation for the overall
design of our study.  We describe the data used and its treatment
in \S~\ref{sec:data}, and in \S~\ref{sec:target} describe the target
selection criteria, justification, and implementation. We discuss sources
of statistical contamination in \S~\ref{sec:contamination}. The most
salient points are synthesized in \S~\ref{sec:discussion} as groundwork
for \citet{trt_p2}, where the helioseismic analysis is presented, and
for \citet{trt_p3}, where the statistical analysis of the helioseismic
and magnetic data are presented.

\section{Study Design}
\label{sec:design}

The goal of this study is to determine whether there
exists a pre-emergence signature of solar active regions visible using
local helioseismic methods and understand said signal, if it exists,
in the context of active-region formation theory.  As summarized above,
case studies have led to conflicting results.  We have designed a study
that utilizes appropriate statistical tests applied to data which include
``control'' samples.  Such a study requires two basic things: sufficient
samples of both ``event'' data and a control set, and care in selecting
both samples so as to minimize bias.

It is fortunate now that there are sufficient data available to perform such a
study, including a statistical analysis of the results.  The statistical method
we use in \citet{trt_p3} is discriminant analysis \citep[e.g.,][]{Kendall1983},
a technique that tests for {\it any} difference between the two samples.  As
such, any systematic bias that is present in the sampling from one population
but absent in the other may appear as a false discriminant.  For example, if
all samples for one population were obtained from east of central meridian while
all samples for the other were obtained from west of central meridian, then
the samples could be differentiated simply due to a bias in the Doppler signal
from solar rotation, not a true detection of emergence.  We refer to this bias
as statistical contamination.

The basic data comprise time-series of Doppler velocity obtained at
the solar surface, from which shifts in subsurface travel-times are
derived using helioseismic holography \citep{Lindsey2000,Braun2007}.
Obtaining a reliable seismic signature requires a temporal sequence
of data, the length of which will govern the signal-to-noise ratios
of the inferred subsurface patterns; yet the data quality may degrade
with proximity to the solar limb.  These realities create limits on the
observable solar disk available for drawing the samples.

In the case of analysis using helioseismology, bias may take
many forms.  Due to the global frequency shifts with solar cycle
\citep{Woodard1985,JCD2002,Chaplin2007} the Doppler velocity signals
may have a component distinctly linked directly to the date.  Systematic
effects \citep{Braun2008,Zhao2012,Baldner2012} may create a dependence
of the helioseismology results on apparent disk position.  Active regions
emerge within a fairly narrow latitude range which itself shifts with the
phase of the activity cycle, leading to another potential source of bias.

To allow an unambiguous detection of subsurface signals, the emergence
episodes should be isolated in time and space from other strong magnetic
sources and nearby emergence episodes.  Yet active regions often emerge
in close proximity to already-established active regions or remnant
fields \citep{PetrovayAbuzeid1991,HarveyZwaan1993,PojogaCudnik2002}.
The controls must ideally also have no magnetic emergence occurring,
and minimal strong-field regions within the immediate field of view, but
they must also match the magnetic context of the population of emerging
targets, as the solar disk gets crowded with active regions and their
remnants during the solar maximum years.

Thus, it is key to couple observations of the solar surface magnetic
field and its evolution to the selection and characterization of the
seismology data.   Pairing the magnetic data to the seismic data provides
guidance for interpreting any seismic signature observed, both in the
control and event groups.

The study is designed based on the following steps:
\begin{enumerate}
\item Locate and identify a statistically significant sample of the population
of new active
region appearances, according to constraints imposed to minimize bias
and noise.
\item Locate and identify a sample of the emergence-free population, matched in 
time and position to the pre-emergence sample,
to serve as a control.
\item Apply helioseismic data analysis ``blindly'' to the two samples.
\item Parametrize the results from the helioseismic analysis
and magnetic field data. 
\item Apply Discriminant Analysis to the seismic and magnetic parameters to quantify 
the differences between the two samples.
\end{enumerate}

\section{Data}
\label{sec:data}

A study such as this requires a statistically significant sample drawn
from the populations in question.  Limitations posed due to observational
and statistical constraints, described in detail below, thus pointed to
using data from the Global Oscillations Network Group (``GONG''), from
the era after the camera upgrades \citep[beginning in 2001,][]{gong,gong1}.
The GONG system records wavelength-modulated full-disk images sampled
at $2.5\arcsec$ for $5\arcsec$ optical resolution, from which Doppler
signals are retrieved on a 1-minute cadence.

Key to interpreting any detected seismic signature is knowing the
``landscape'' of the surface magnetic field.  As we are specifically
interested in pre-emergence signatures, the surface magnetic fields
and the signature of magnetic flux emergence {\it define} the timing
for the entire project.  At the time of design and implementation of this study,
the line-of-sight field from the GONG data were not readily available.
We thus rely upon the full-disk line-of-sight component magnetic
field data from the Michelson Doppler Imager aboard the Solar and
Heliospheric Observatory \citep[SOHO/MDI,][]{mdi}.  Specifically, we used the level
1.8.2 synoptic data acquired with a 96-minute cadence and $1.98\arcsec$
pixel size \footnote{Emergence times were initially determined
using earlier level 1.8.1 data, but we do not expect any systematic
differences as the emergence times were based on the change of the signal,
not a pre-determined threshold.} to qualitatively and quantitatively
evaluate the magnetic landscape of the samples.

Helioseismic data from MDI were not used in this study for two reasons.
First, the high-rate full disk data (``dynamic campaigns'') are only available for a few months per year,
limiting the data available for a statistical study.
Second, the medium-$\ell$ (``structure'') data are not optimal for studying wave
propagation at distances less than approximately ten heliocentric degrees \citep{Giles2000}
whereas the present work examines depths $\la 25\,{\rm Mm}$ which requires small 
distances.
For these reasons, we have used the GONG data for the helioseismic analysis
performed in this study. 

\subsection{Target Selection Criteria: The ``PE''s: Pre-Emergence Regions}
\label{sec:target}

The initial target list for emerging active regions was derived
from the ``Sunspot Group Reports'' produced by USAF/NOAA and
available through the National Geophysical Data Center\footnote{\tt
http://www.ngdc.noaa.gov/stp/solar/sunspotregionsdata.html}.  The date-range
used was chosen according to requirements for the helioseismology data, and covered
July 2001 -- November 2007.  Regions listed as first appearing within $\theta
\leq 30^\circ$ of disk center and which achieved an area $>
10\times 10^{-6}$ hemispheres ($\mu{\rm H}$) during their disk passage
determined the initial target list, and the initial emergence times
and locations, that were subsequently refined.

\begin{figure}
\plotone{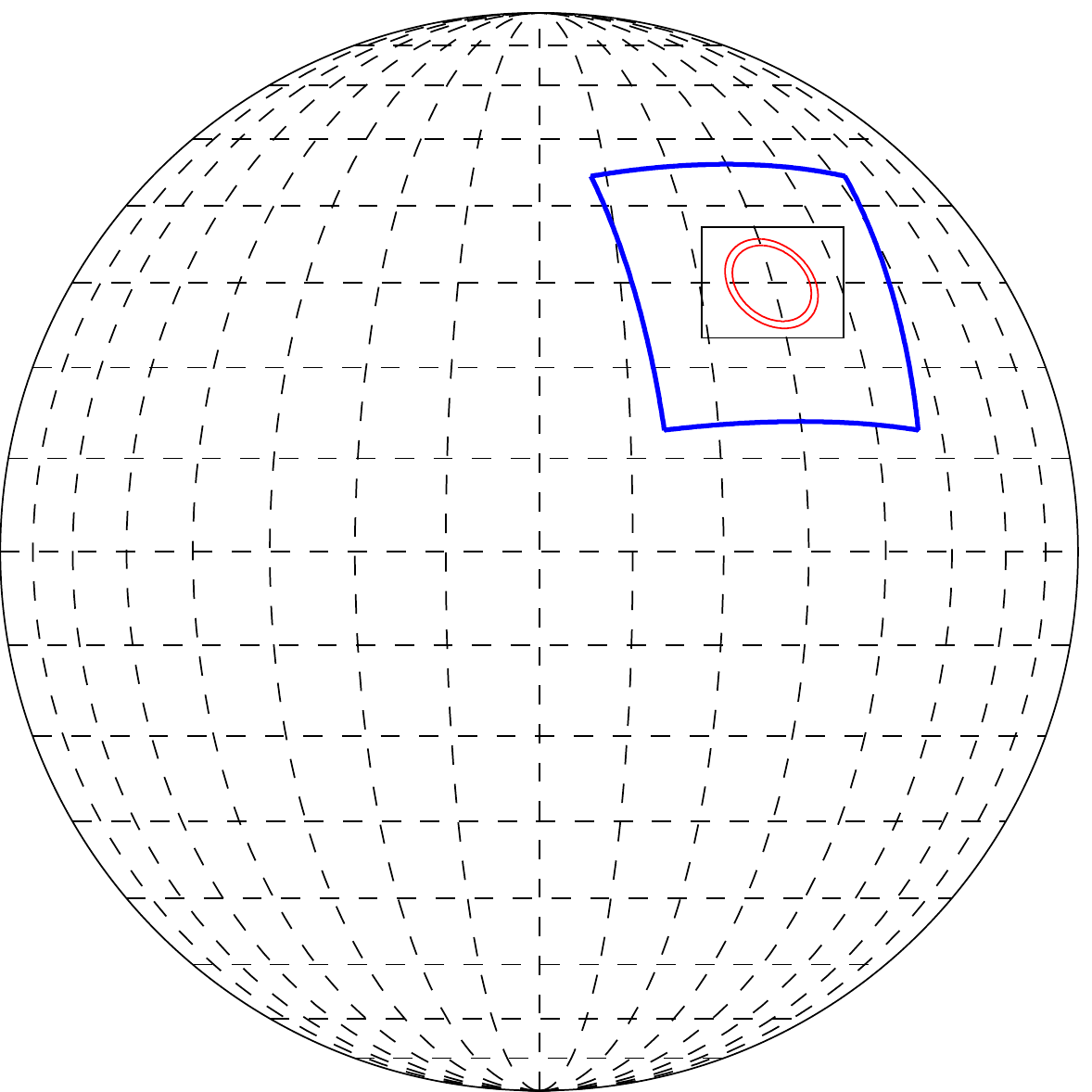}
\caption{A schematic showing the relative sizes of areas considered
during the data preparation and analysis.  The solar Stonyhurst disk is shown
with lines at $10^\circ$ latitude and longitude intervals (- - -).  The black box
indicates $128\times 100$-pixel area of an MDI image used
for the initial evaluation, in this case centered at N30 W30.
The larger box (blue) is a Postel projection region $32^\circ \times 32^\circ$,
showing the area extracted for the tracked Doppler data from GONG, and the
corresponding area of computed radial-component of the field 
from MDI extracted for the full analysis.  The red circles indicate
the size and width of the largest annulus (filter ``TD11'') used for 
computing helioseismology parameters \citep[see][]{trt_p2}.}
\label{fig:cartoon}
\end{figure}

\begin{figure}
\centerline{
\includegraphics[width=0.3\linewidth]{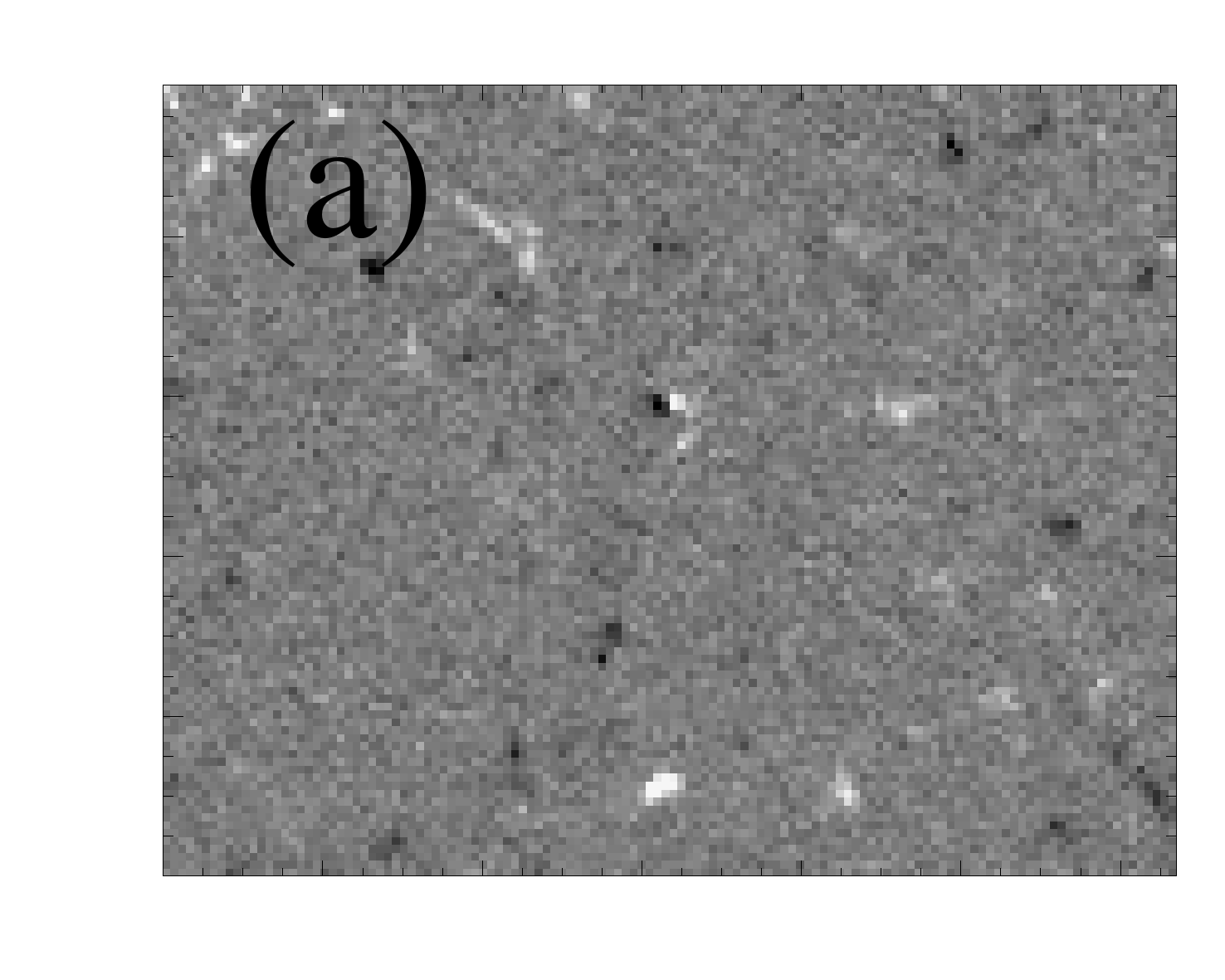}
\includegraphics[width=0.3\linewidth]{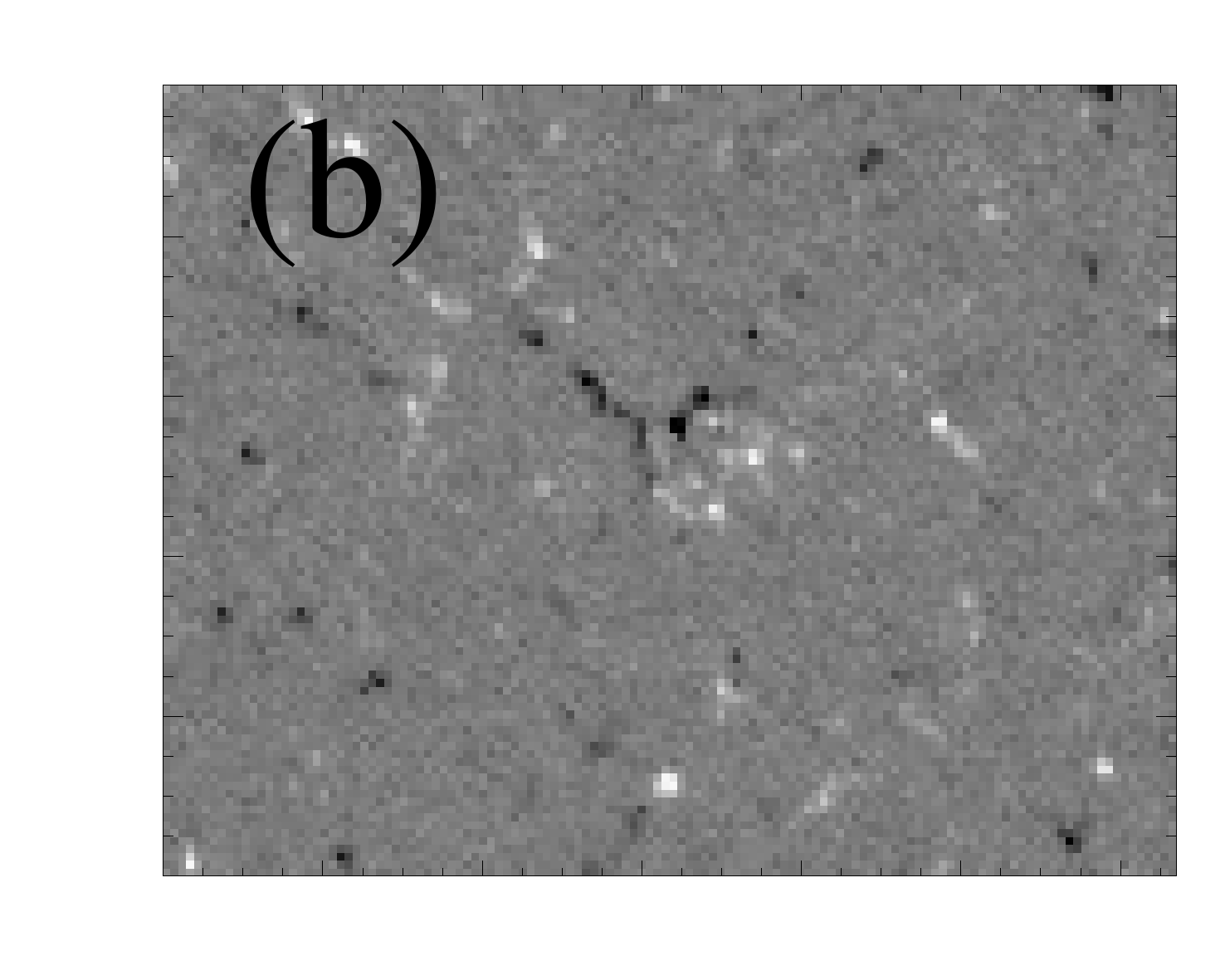}
\includegraphics[width=0.3\linewidth]{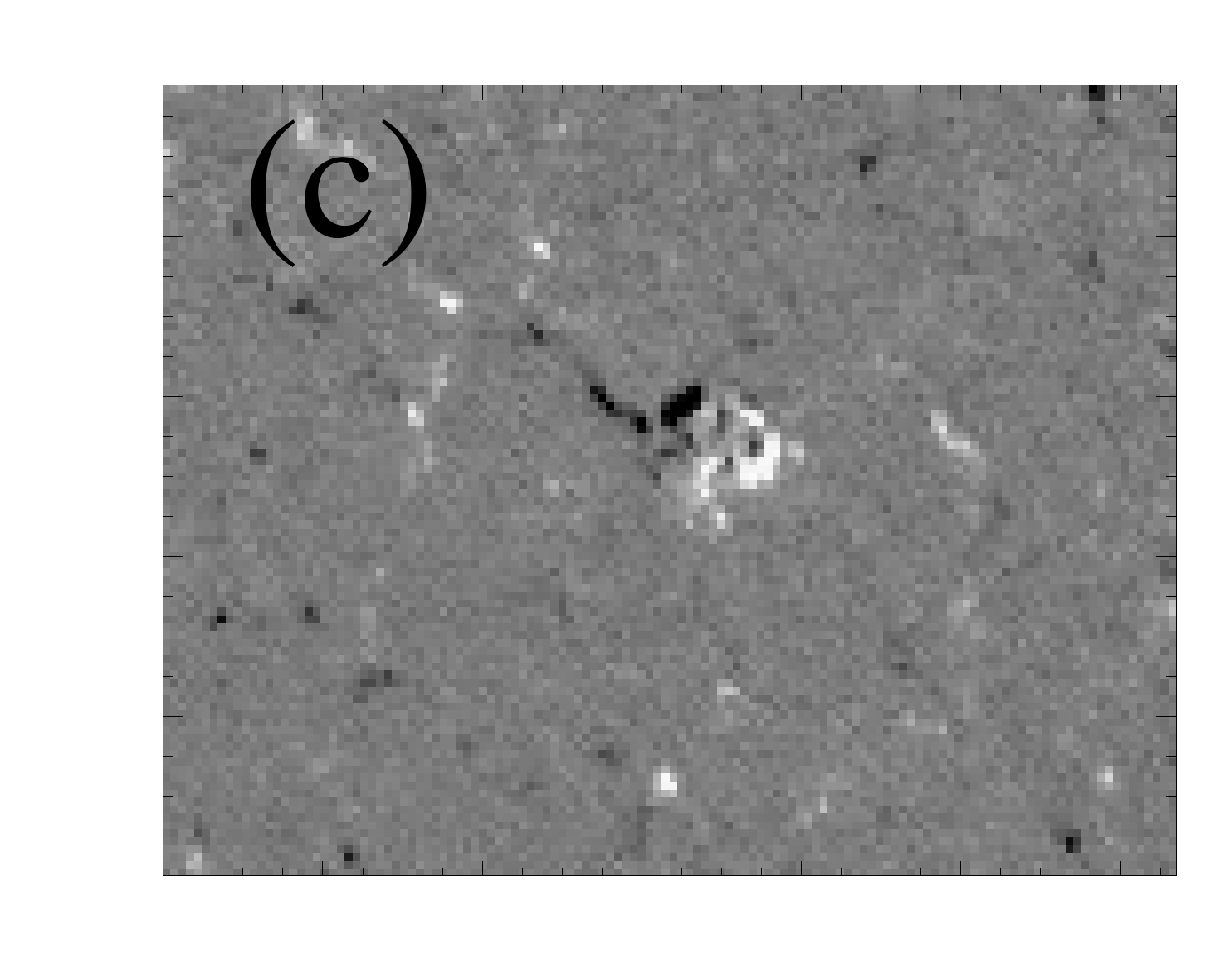}}
\centerline{
\includegraphics[width=0.3\linewidth]{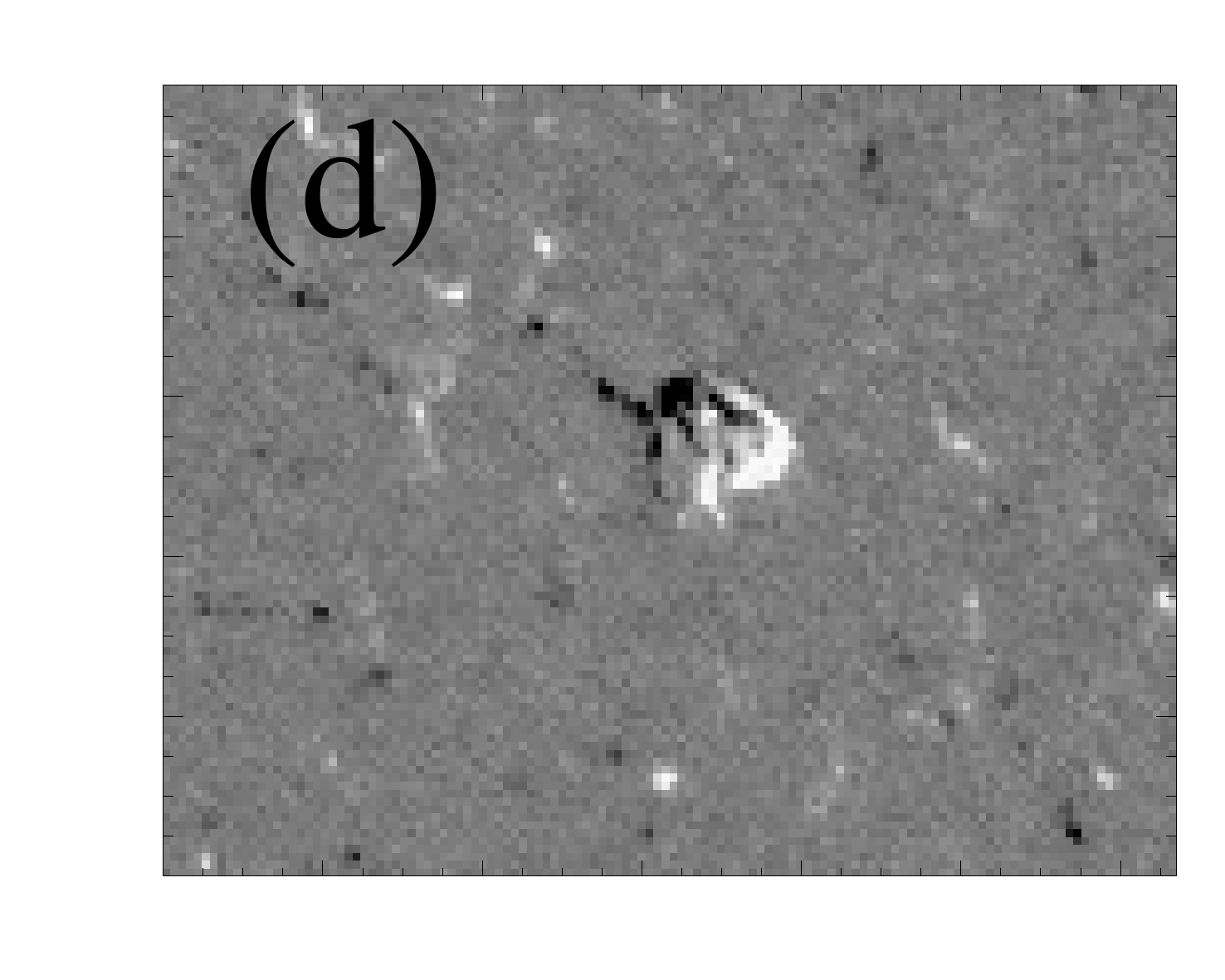}
\includegraphics[width=0.3\linewidth]{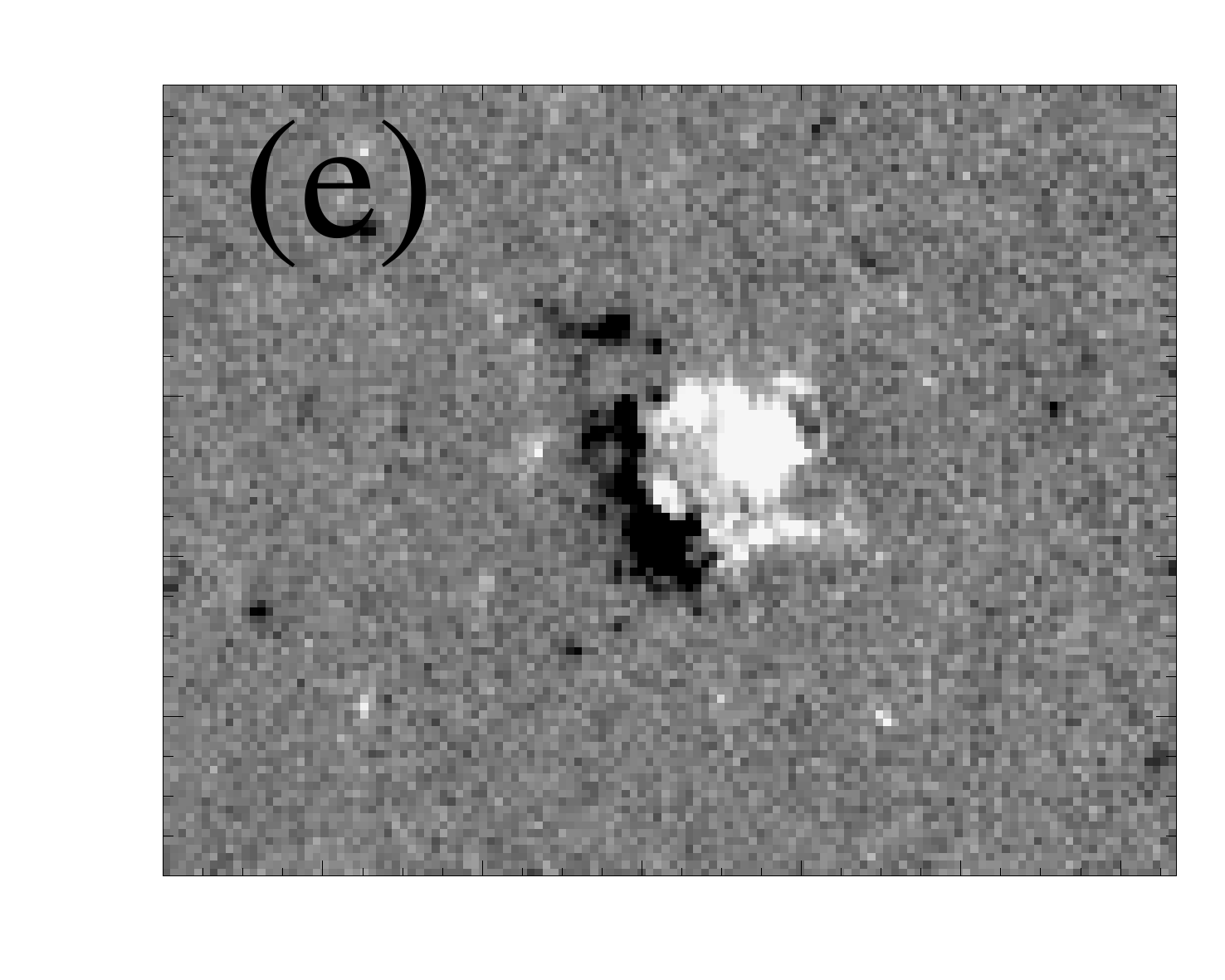}
\includegraphics[width=0.4\linewidth]{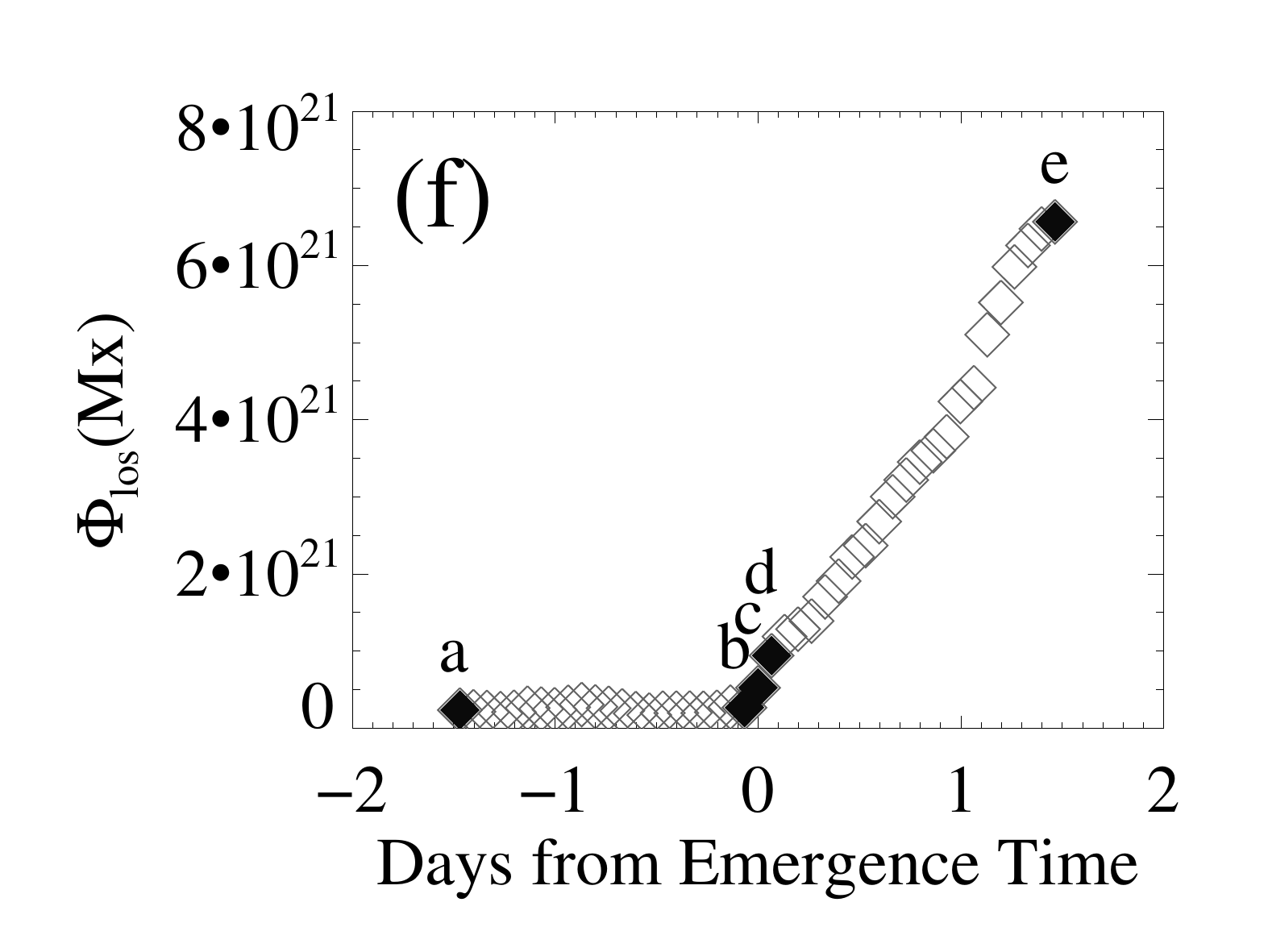}}
\caption{An example of a Pre-Emergence target, 
NOAA AR 10559 which had an assigned emergence time $t_0$ of 2004-02-13T11:15:02.677Z 
(see text for details) at N07 W22.4.  The images (a-e) are the $128\times100$-pixel
images from the MDI full-disk line-of-sight magnetic data used for initial
evaluation of the emergence episode, all scaled to $\pm 500$\,G.
The image (c) shows the assigned ``emergence time'' $t_0$.
The temporal evolution of the pseudo-flux $\Phi_{\rm los}=\sum|\Bl|/\mu\,\Delta{\rm A}$ for 
this test field of view is shown in (f), as a function of time relative to
the inferred time of emergence, determined as the first MDI magnetogram
when 10\% of the eventual maximum change in flux, $\Phi_{\rm los}$, has appeared (c).
Data points for the images shown are filled in and labeled; a mix of 
30\,s and 300\,s MDI data are both used and shown here, evident by the
different apparent noise levels.
}
\label{fig:t0imagesPE}
\end{figure}

MDI 3-day time-series were constructed centered on this initial emergence
date and time, using a fixed $128 \times 100$ pixel box centered on the
initial emergence location (see Figure~\ref{fig:cartoon} for a schematic),
and tracked with the synodic rotation rate (Figure~\ref{fig:t0imagesPE}).
As a check against extreme viewing angles at the beginning or end of the
time-series, additional limits on the edges of the box were placed at
E41 and W67 heliographic longitude (East longitudes are $<0)$) and $\pm
60^\circ$ heliographic latitude.  The $\Bl$ data were initially summed
to a pseudo-``flux'', $\Phi_{\rm los} = \sum|\Bl|/\mu\,\Delta{\rm A}$,
where $\mu = \cos{\theta}$ and $\theta$ is the observing angle, and
$\Delta{\rm A}$ is the physical area of a pixel.  A refined emergence
time, $t_0$, was defined as the time of the first MDI observation
after $\Phi_{\rm los}$ reached 10\% of the maximum achieved (minus any 
flux present at the beginning of the time-series) over the
time series.  That is, the ``10\% rule'' refers to 10\% of the maximum
increase detected.  The kurtosis (fourth moment) of the distribution of $\Bl$
in the frame generally increases dramatically at the time of emergence,
signifying a distinct change in the spatial distribution of $\Bl$;
a sudden change in the kurtosis was used to confirm the ``10\% rule''
but was not relied upon in isolation.  Thus, the emergence time is only
defined within the 96-minute MDI cadence.  For the analysis methods later
applied, which require many hours of data, there is little to be gained
by refining this definition further.  The NOAA reports of active region
coordinates were generally accurate, although our definition of $t_0$
was generally earlier than the NOAA reports by anywhere from a few hours
up to a day.

\begin{figure}
\centerline{
\includegraphics[width=0.3\linewidth]{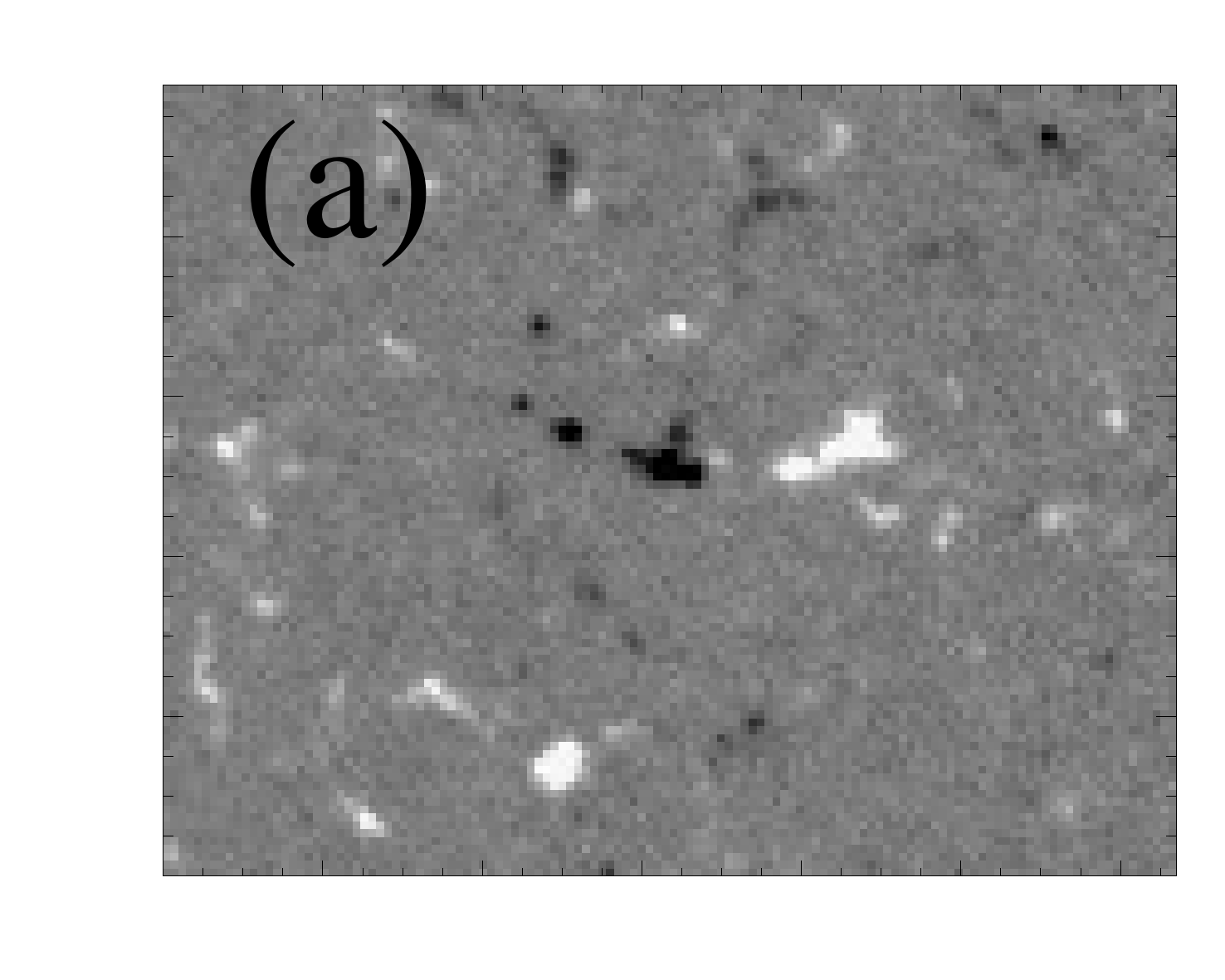}
\includegraphics[width=0.3\linewidth]{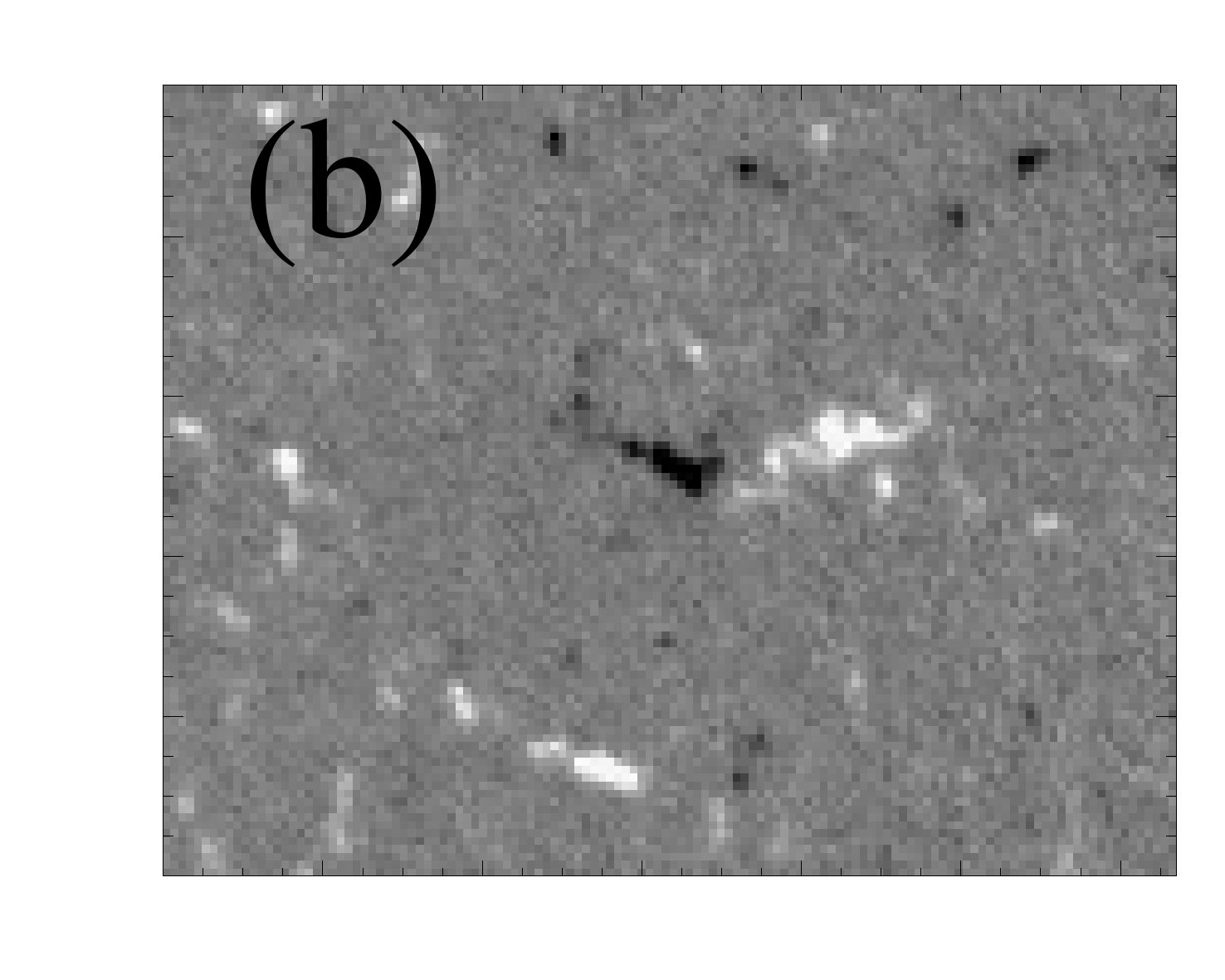}
\includegraphics[width=0.3\linewidth]{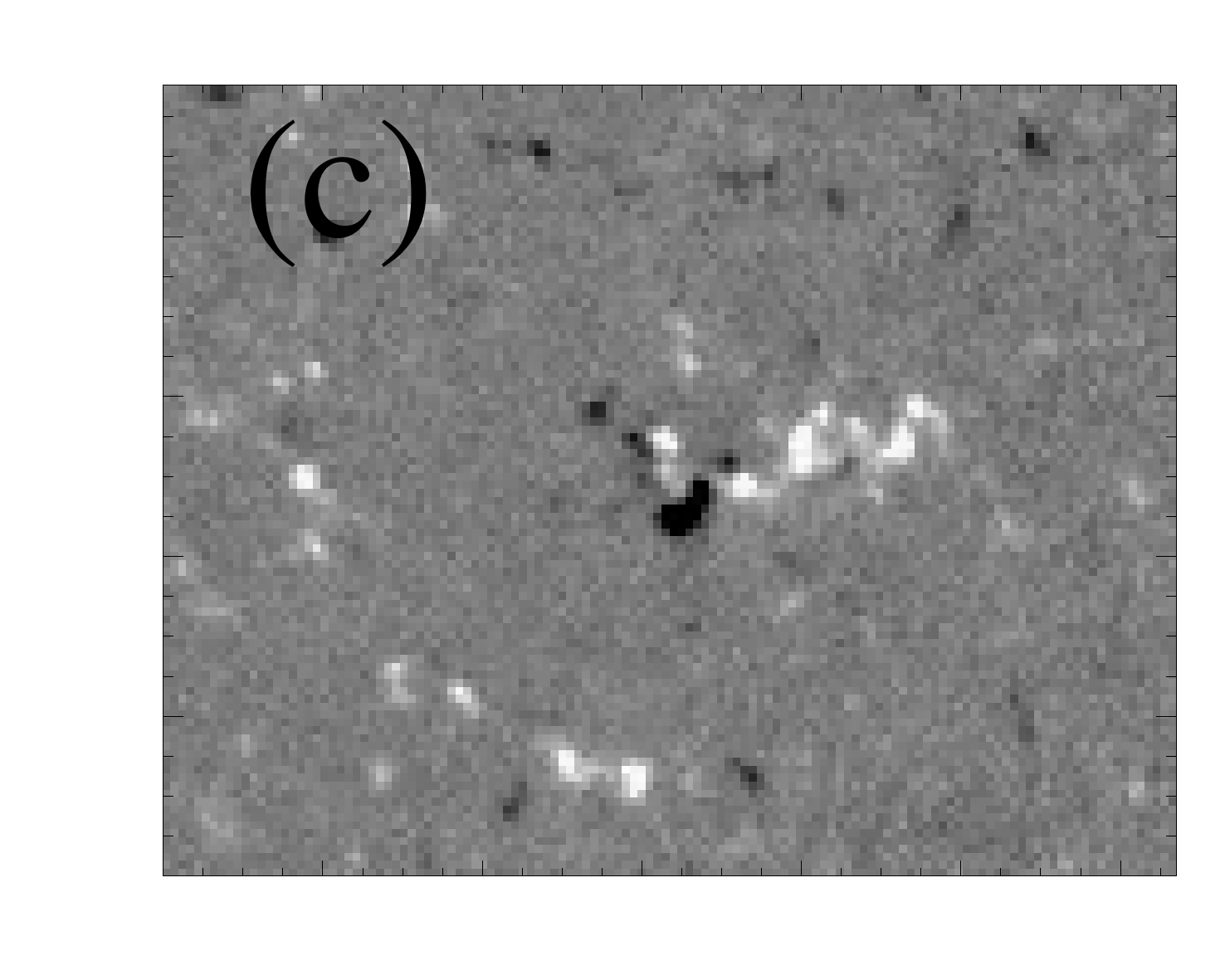}}
\centerline{
\includegraphics[width=0.3\linewidth]{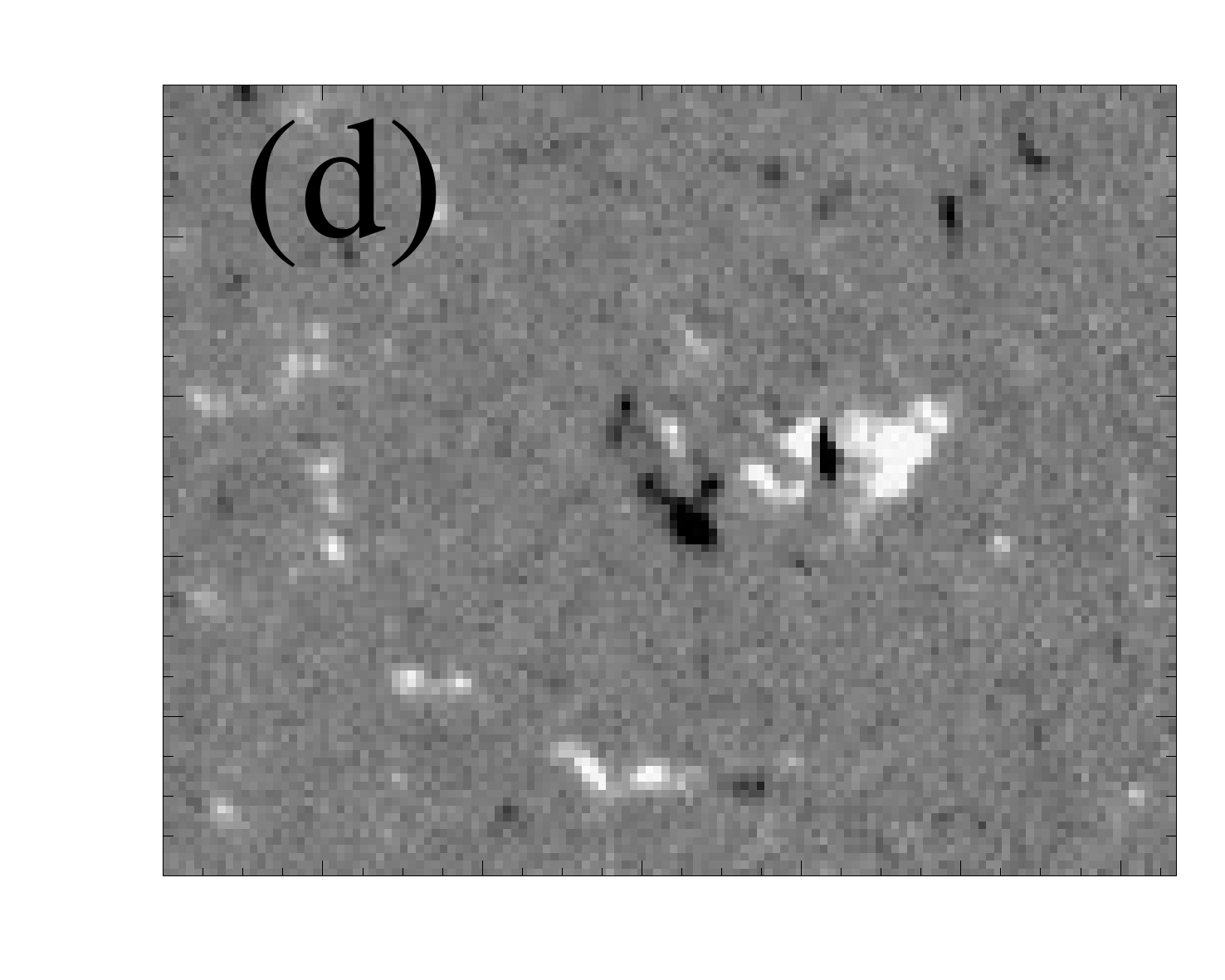}
\includegraphics[width=0.3\linewidth]{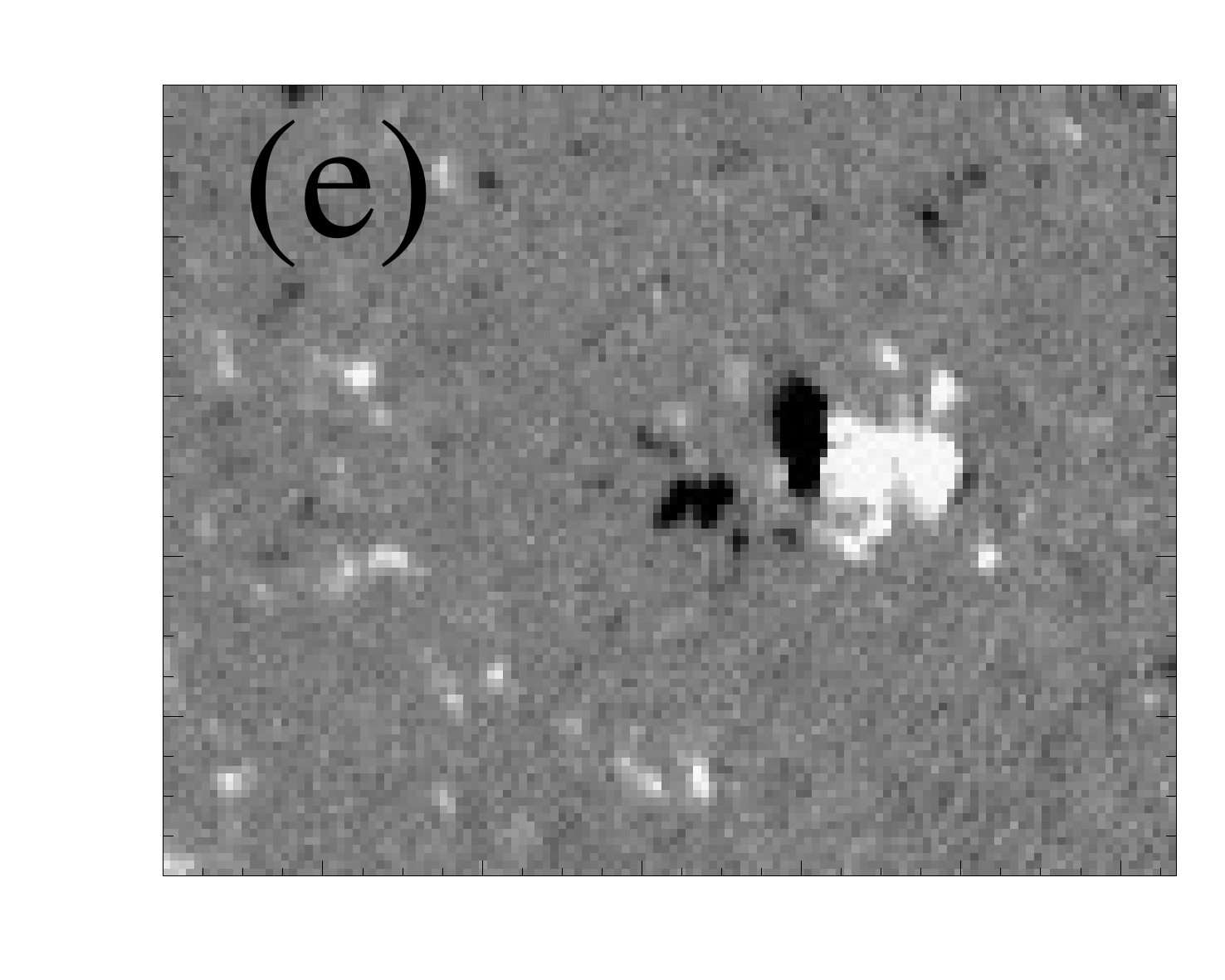}
\includegraphics[width=0.4\linewidth]{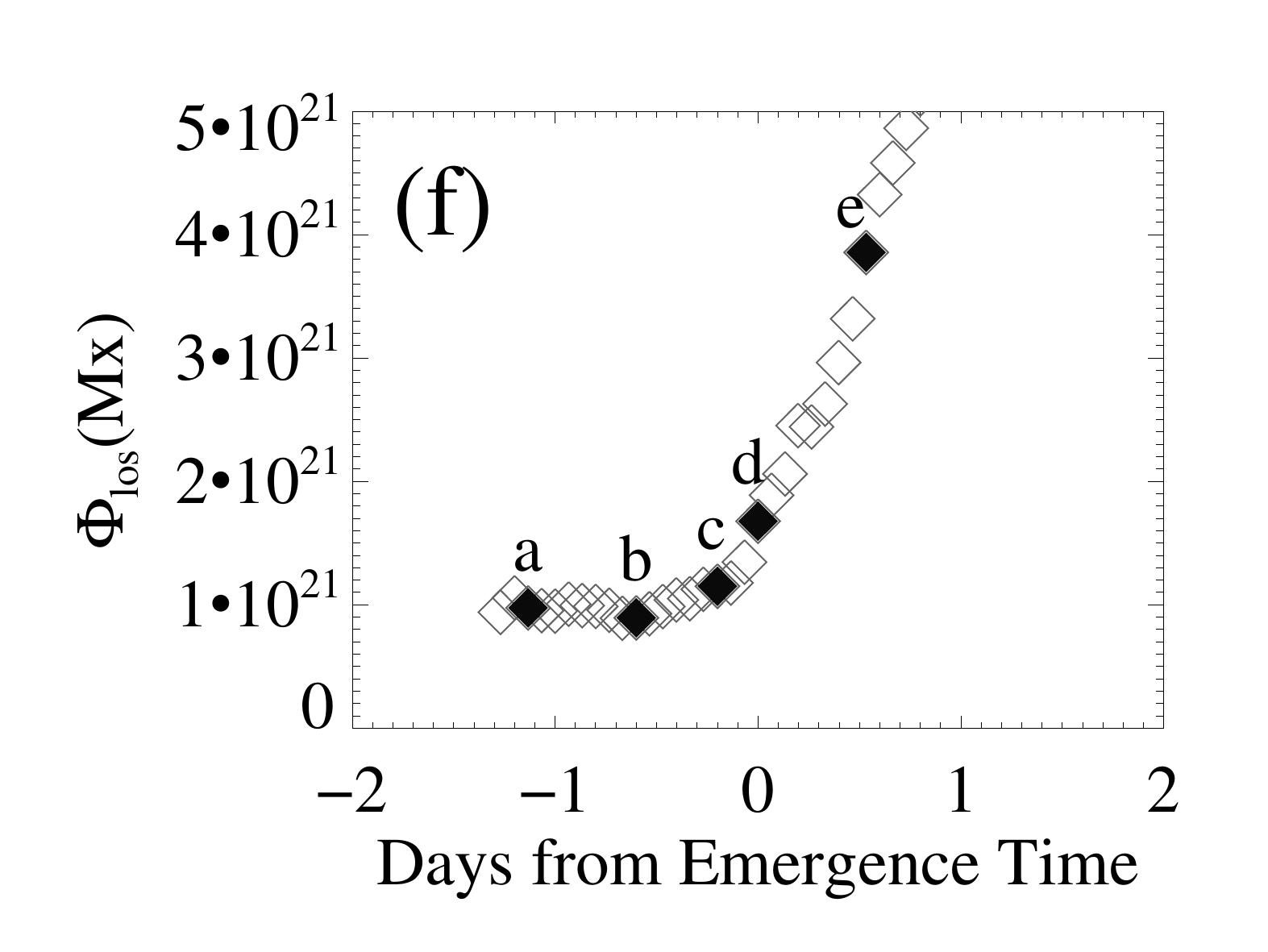}}
\caption{An example of a Pre-Emergence target with a less-clear emergence time, 
NOAA AR 9564 which had an assigned emergence time $t_0$ of 2001-08-01T06:27:01.250Z
(see text for details) at N14 W21.2.  The images (a-e) are in the same
format as in Figure~\ref{fig:t0imagesPE}.
The temporal evolution of $\Phi_{\rm los}$ 
is shown in (f), except that the maximum of the region attained is truncated
to better show the early evolution.  Data points for the images shown are filled in and labeled.
In this case, the ``background'' 
$\Phi_{\rm los}=0.96\times 10^{21}$Mx is larger than the previous example.
The region eventually reached $9.5\times 10^{21}$Mx, or a maximum
increase of $8.6\times 10^{21}$Mx, hence point (d) at $1.8\times10^{21}$Mx
was the identified emergence time by the objective algorithm.
However, it is clear that a small episode of flux emergence
apparently occurred between (b) and (c) as well.  While in this case we can
argue a 6hr uncertainty in the emergence times, there was very little if any surface signal
of the emergence for many hours prior to (c).
}
\label{fig:t0imagesPEmess}
\end{figure}

Emergence of surface field is rarely a smoothly monotonic process
\citep{Zwaan1985,roadmap,kubo_etal_2003}.  An example of that reality
is shown in Figure~\ref{fig:t0imagesPEmess} (and discussed further
in Section~\ref{sec:mess}).  As such, the flux history and thresholds
here constitute a selection rule to be used for a statistical approach,
rather than a profound statement of solar physics.  And as such, there
will be regions for which the definition blatantly misses the mark of
rising flux presence.  The goal here is a well-defined ``good option'',
that is objective and repeatable for a statistically-significant sample
of data.

Regions were rejected for a number of reasons, primarily data-gaps
(in either MDI at or near the emergence time, or GONG data for final
analysis) or immediate proximity (within the $128 \times 100$ pixel
box) of another active region.  No further tests were made concerning
the eventual size of the active region or speed of emergence; a later
subjective evaluation rejected regions if $t_0$ appeared incorrect by
more than a few MDI-derived data points.  The fixed box used at this
stage was fairly restrictive.

\begin{figure}
\plotone{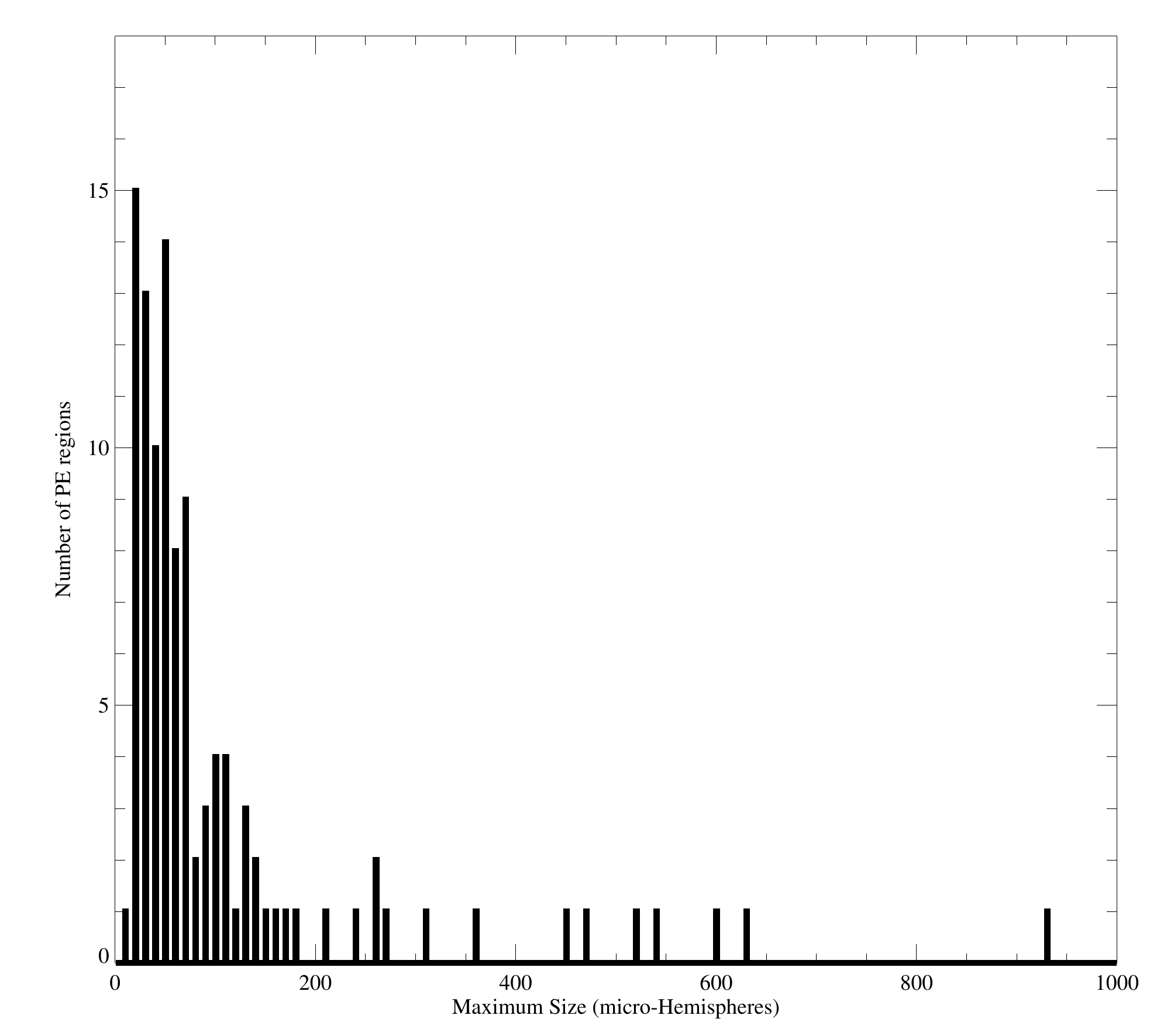}
\caption{Histogram of the maximum size of the sunspot group attained during disk visibility 
of the emerging active regions, in $\mu$H (micro-hemispheres) as reported
by the NOAA active region lists. The minimum reported size is 10$\mu$H; the 
largest included in this sample was 930$\mu$H.}
\label{fig:maxsizehisto}
\end{figure}

The refined location and time of emergence, defined as above, were
used to generate the Doppler-velocity data (see \S~\ref{sec:gongprep}).
The final result is 107 pre-emergence (``PE'') target regions between
2001 and 2007.  In Table~\ref{table:PE} we list the identifying features
of these regions: the NOAA Active Region number, the $t_0$ as defined
above, and the latitude and longitude of the center of the $128 \times
100$ pixel box at that time.  Note that the longitude was generally
refined from the NOAA reports, while the latitude generally was not,
and as such is effectively an integer.  In Figure~\ref{fig:maxsizehisto}
we show the final distribution of the (eventual) maximum size achieved 
(as reported in the NOAA compilations) for the active regions in the PE list.

A subset of eleven regions are singled out as being particularly ``clean'',
and these are indicated with a superindex ``a'' in Table~\ref{table:PE}.  The criteria for this
list are completely subjective: no neighboring active region in the
extracted areas, a very flat pre-emergence flux history, and an emergence
characterized by a very uniform and steep slope of $d\Phi_{\rm los}/dt$.
The example shown in Figure~\ref{fig:t0imagesPE} is one such member of
the ``Ultra-Clean Subset''.

\subsection{Target Selection Criteria: The ``NE''s: No-Emergence Control Regions}
\label{section:quiets}

The active-region emergence targets required an accompanying set of
``control data''.  As our final analysis is a statistical analysis based
on the results of both helioseismology-derived and magnetic-derived
parameters, the control data needed to be constructed so as to not
introduce statistical bias into the final distributions.  We outline
the construction of this data set here.

\begin{figure}
\plotone{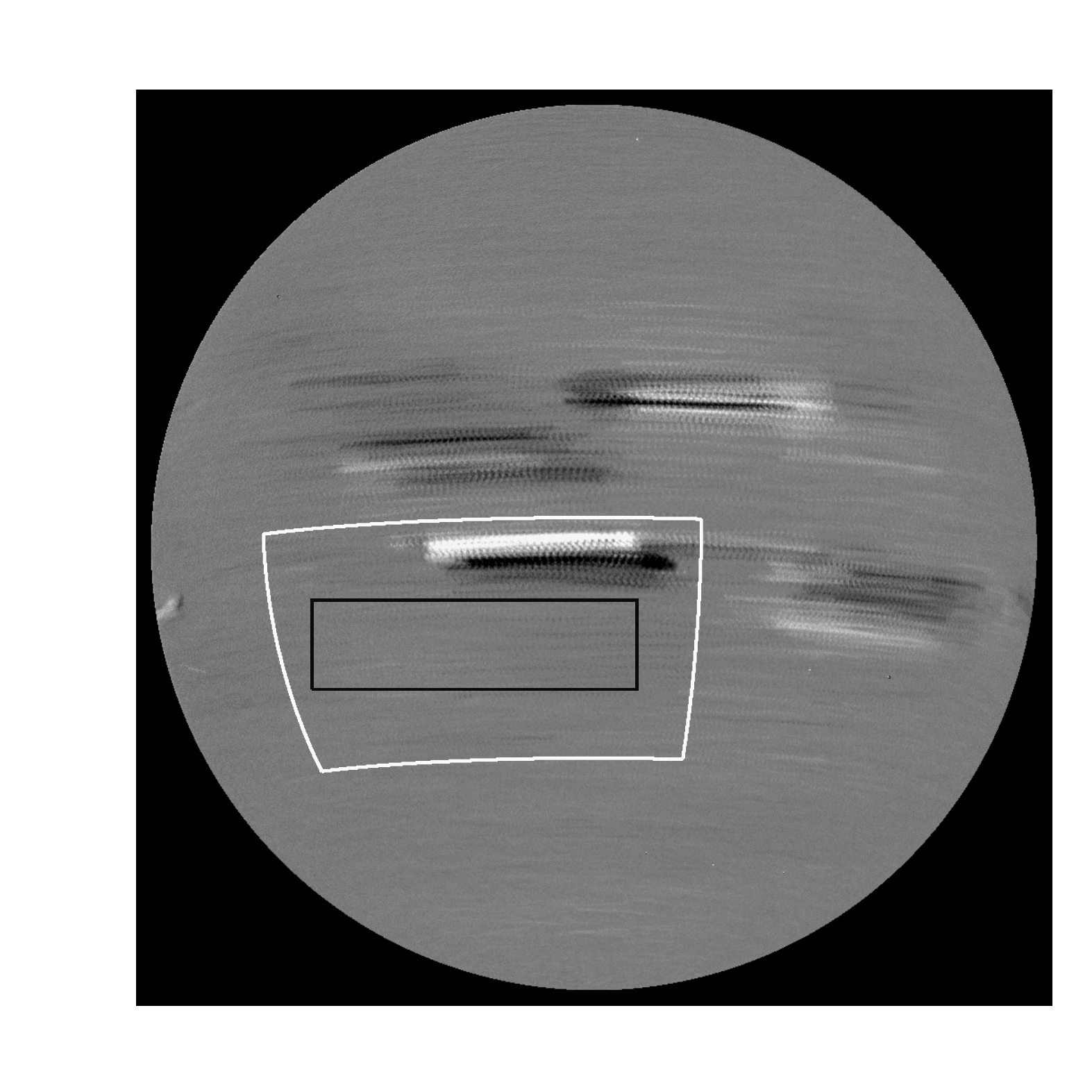}
\caption{The selection area and eventual data-extraction area of a non-emergence
target.  Three days of MDI 96-minute data beginning with MDI orbit \#4053 
(2004-02-06T00:03:02.469Z) have been averaged together, and shown here
scaled to $\pm 100$\,G.  The black box shows the coverage of a $128\times100$-pixel 
box tracked over the three days, 
indicating the entire quiet or ``NonEmerging'' (``NE'') area that consistently
has only signal $< 1000$\,G over the three days.  The white box indicates the area of
tracked GONG and MDI data eventually used for the full analysis, discussed
in Sections~\ref{sec:gongprep},\ref{sec:mdiprep}.
}
\label{fig:nepatch}
\end{figure}

\begin{figure}
\centerline{
\includegraphics[width=0.3\linewidth]{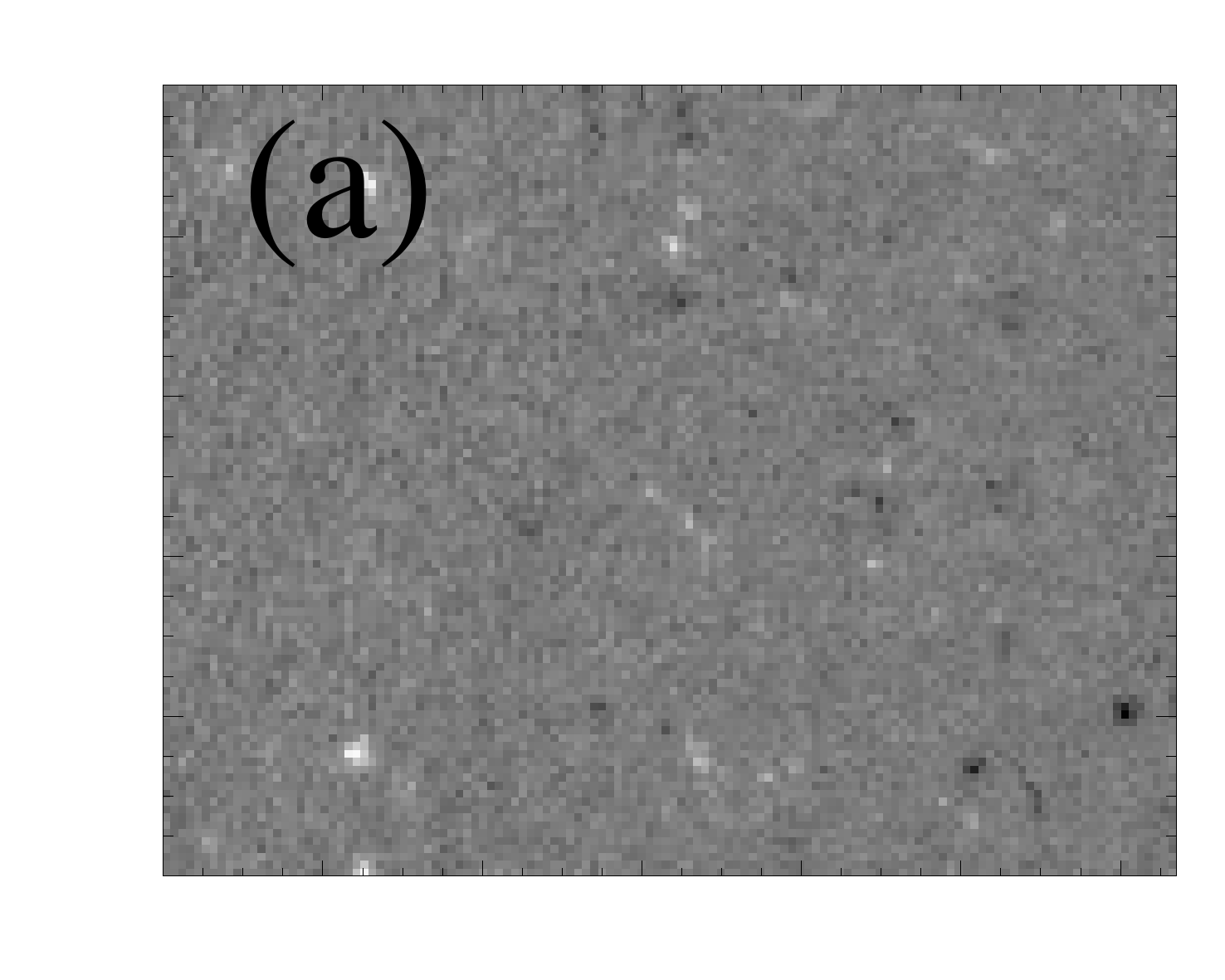}
\includegraphics[width=0.3\linewidth]{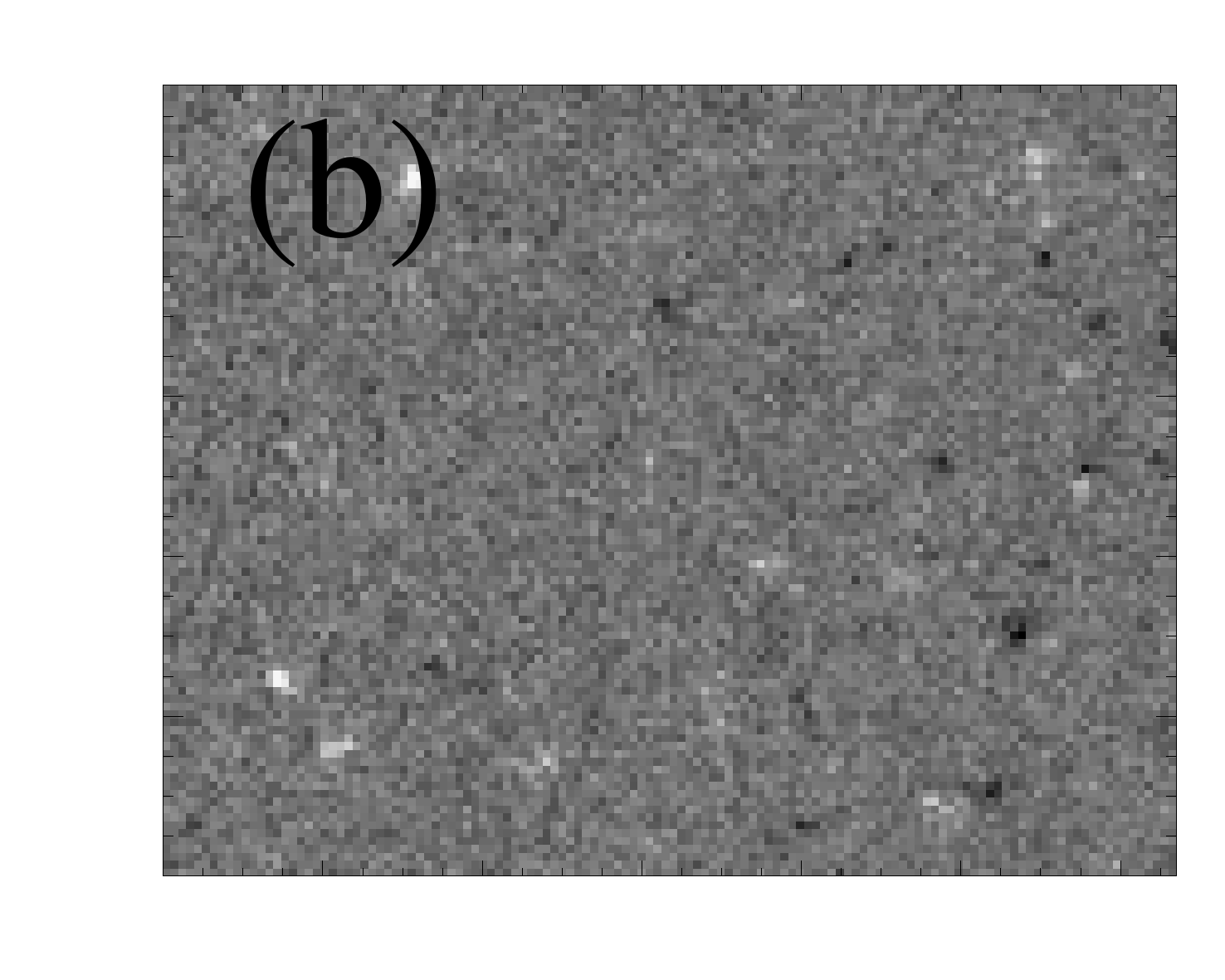}
\includegraphics[width=0.3\linewidth]{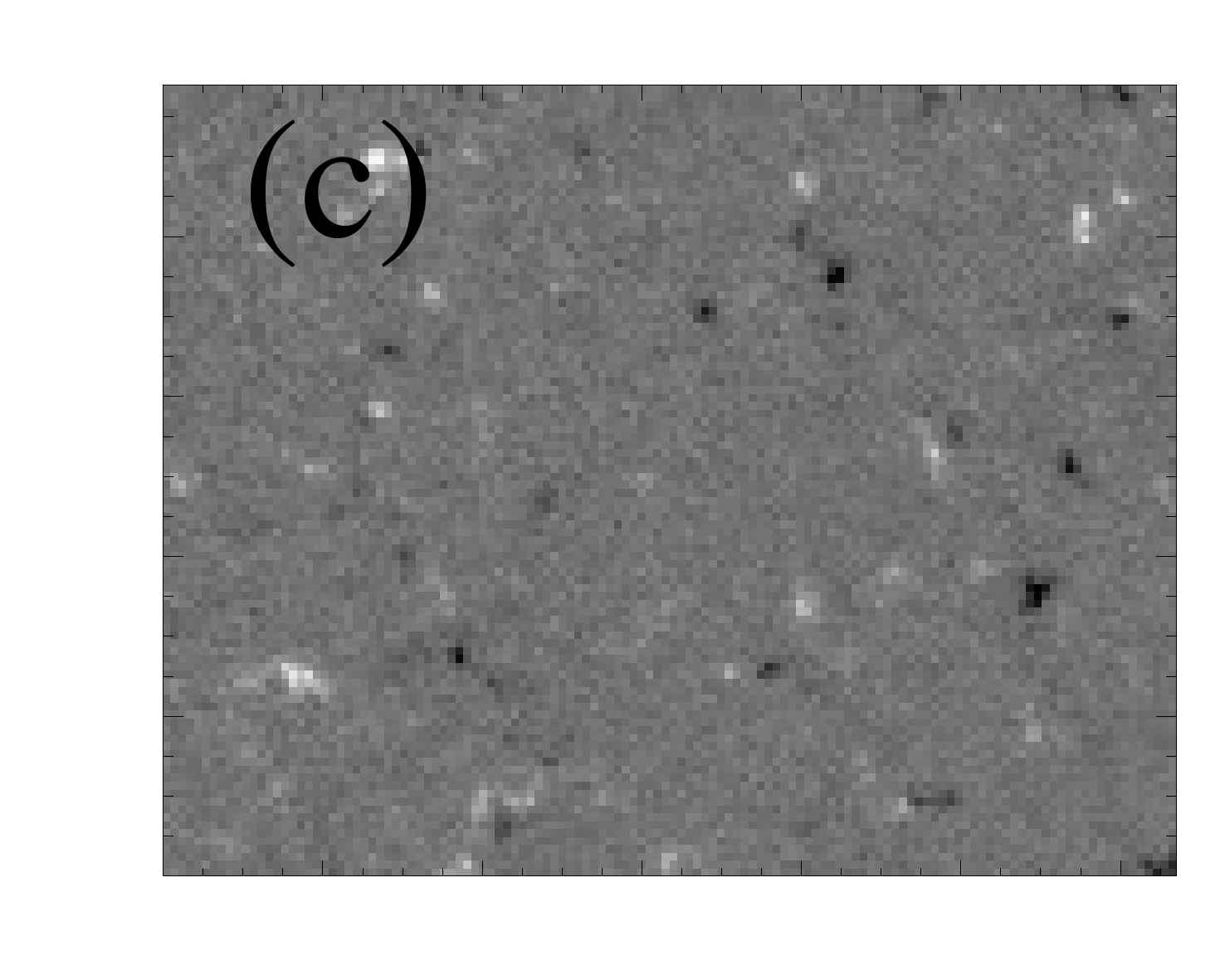}}
\centerline{
\includegraphics[width=0.3\linewidth]{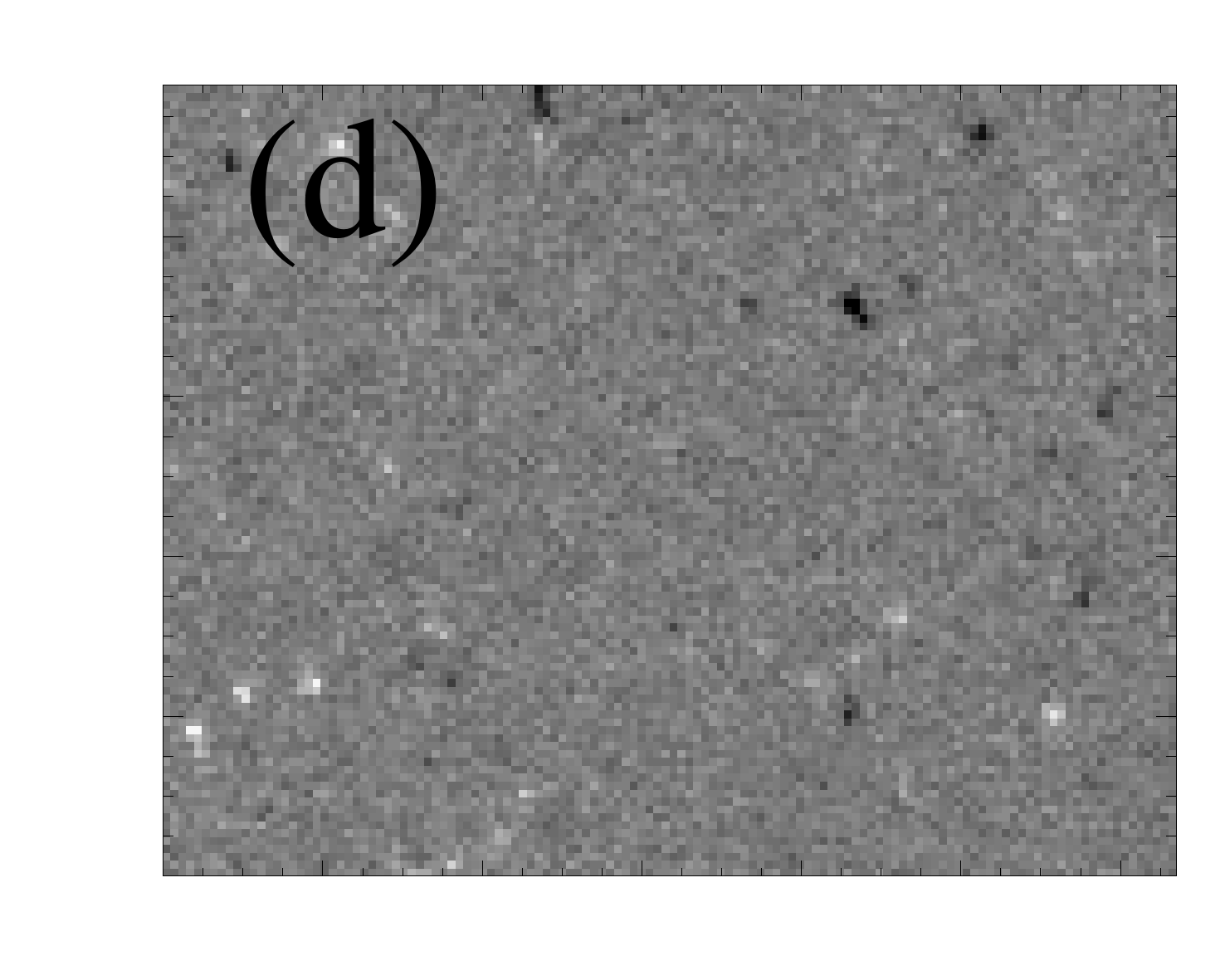}
\includegraphics[width=0.3\linewidth]{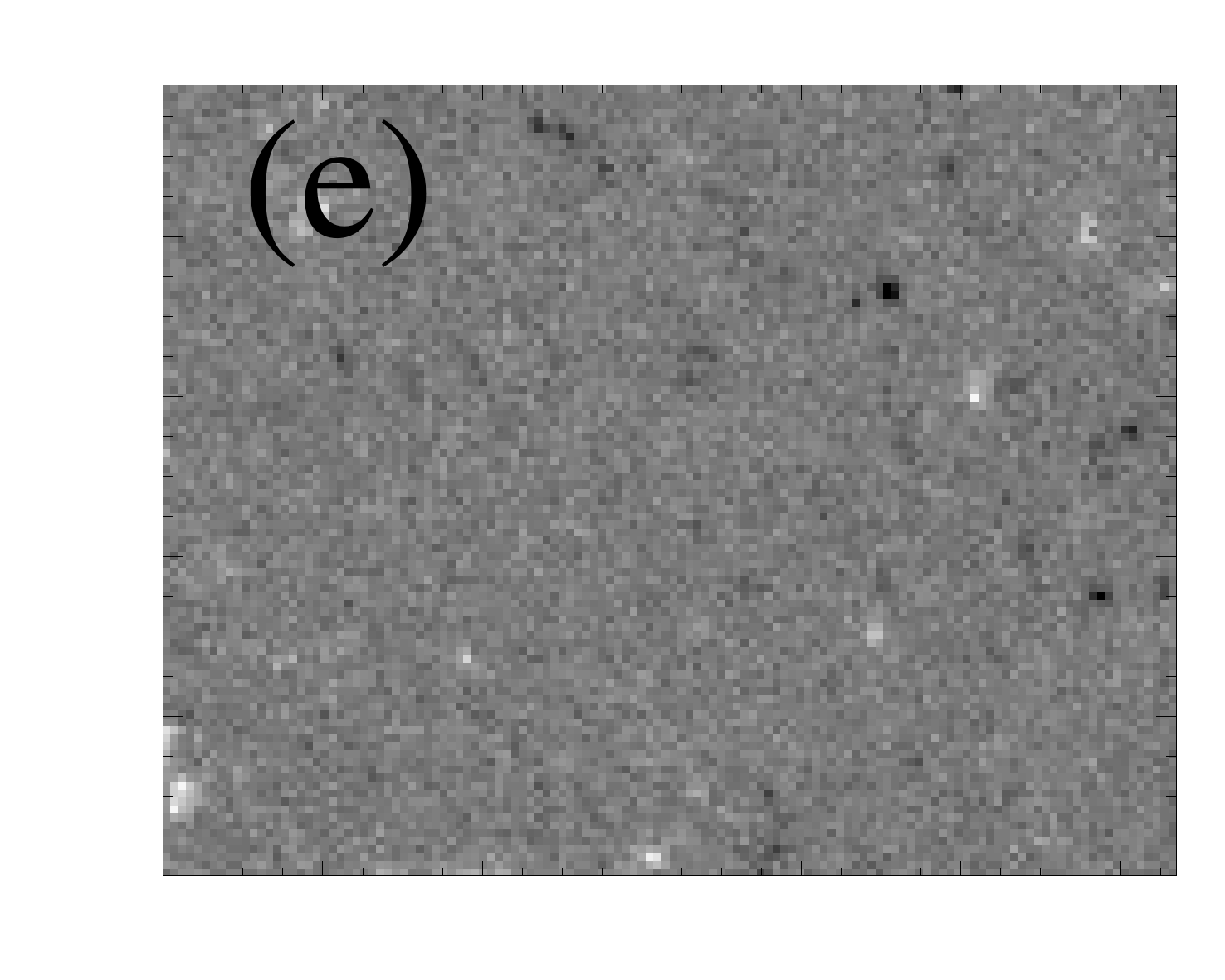}
\includegraphics[width=0.4\linewidth]{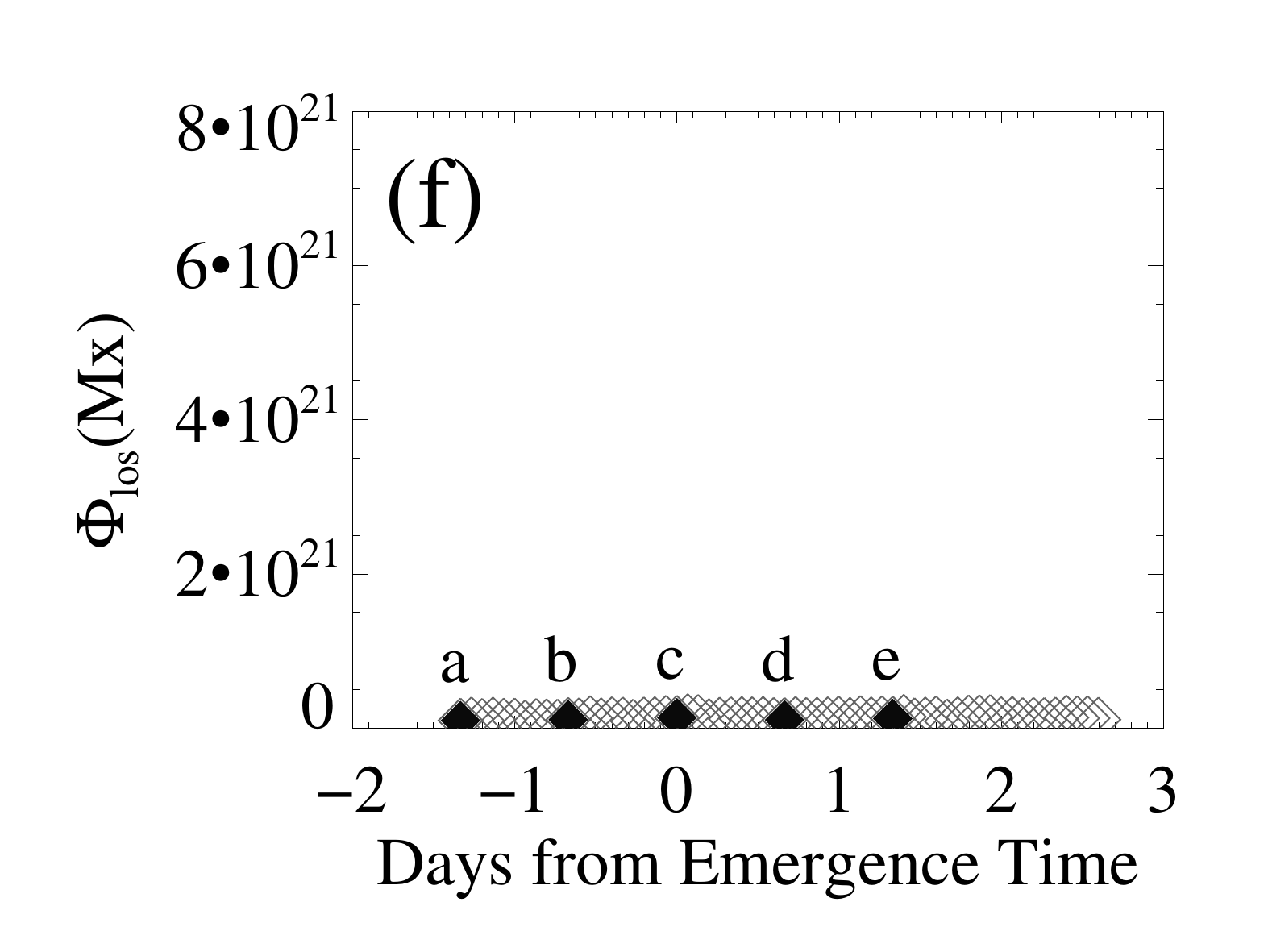}}
\caption{The Non-Emergence target from Figure~\ref{fig:nepatch}
which had an assigned center-time $t_0$ of 2004-02-07T08:03:02.501Z 
(see text for details) at S18.6 E17.4.  The images (a-e) are the $128\times 100$
images from the MDI full-disk line-of-sight magnetic data used for initial
evaluation of the emergence episode, all scaled to $\pm 500$\,G.
The temporal evolution (f) of the pseudo-flux $\sum|\Bl|/\mu\,\Delta{\rm A}$ 
for this test field of view, as a function of time relative to
the inferred time of emergence, scaled to match Figure~\ref{fig:t0imagesPE}.
Data points for the images shown are filled in and labeled.}
\label{fig:t0imagesNE}
\end{figure}

Starting every two MDI days during the same 2001-2007 interval, using the
same-sized $128\times100$-pixel tracked boxes, areas were identified where
the underlying signal stayed consistently $< 1000$\,G\footnote{Gauss are
used as units, with the understanding it is a pixel-averaged quantity.}.
This was accomplished by ``stacking'' three days' worth of MDI data
and extending the target box to effectively cover the tracked area,
as shown in Figure~\ref{fig:nepatch}.  Random locations for these
low-field areas were chosen on the disk for each stack, subject to the
same general constraints as the PE targets with regards to limits on
latitude and longitude.  A time close to the center of the 3-day interval,
falling on an MDI observed time, is designated $t_0$ for the NE data.
While there was the possibility of overlapping areas being chosen,
any randomly-selected NE patch which did overlap was ``weeded out''
as described below.  An example of a ``no-emergence'' region is shown
in Figure~\ref{fig:t0imagesNE}.

This selection algorithm initially provided thousands of possible NE targets
over the seven years.  A subset of approximately 500, selected to generally
follow the distribution in latitude, longitude, and time as the initial set of
PE targets, were used to acquire GONG data (see \S~\ref{sec:gongprep}).  

From these, a subjective evaluation was made, removing approximately
20 targets from consideration primarily due to the existence of small
(obviously, un-numbered) emerging flux regions at the center of the
field-of-view which were not previously detected.  While no specific
criteria were used regarding increasing or changing total flux over the
time interval, the single criterion specified above effectively performed
to constrain selection to regions with impressively consistent magnetic
flux levels, on the whole.

Candidate NE regions were further evaluated and removed if the central
$16^\circ\times16^\circ$ (used for the majority of the helioseismology
analysis, see Section~\ref{sec:gongprep}) overlapped with the central
$16^\circ\times16^\circ$ portion of a PE or another NE at any time.
The final number of NE controls available for distribution control
(see \S~\ref{sec:distribution}, below) was 308.  

\subsection{Distribution Control}
\label{sec:distribution}

\begin{figure}
\plotone{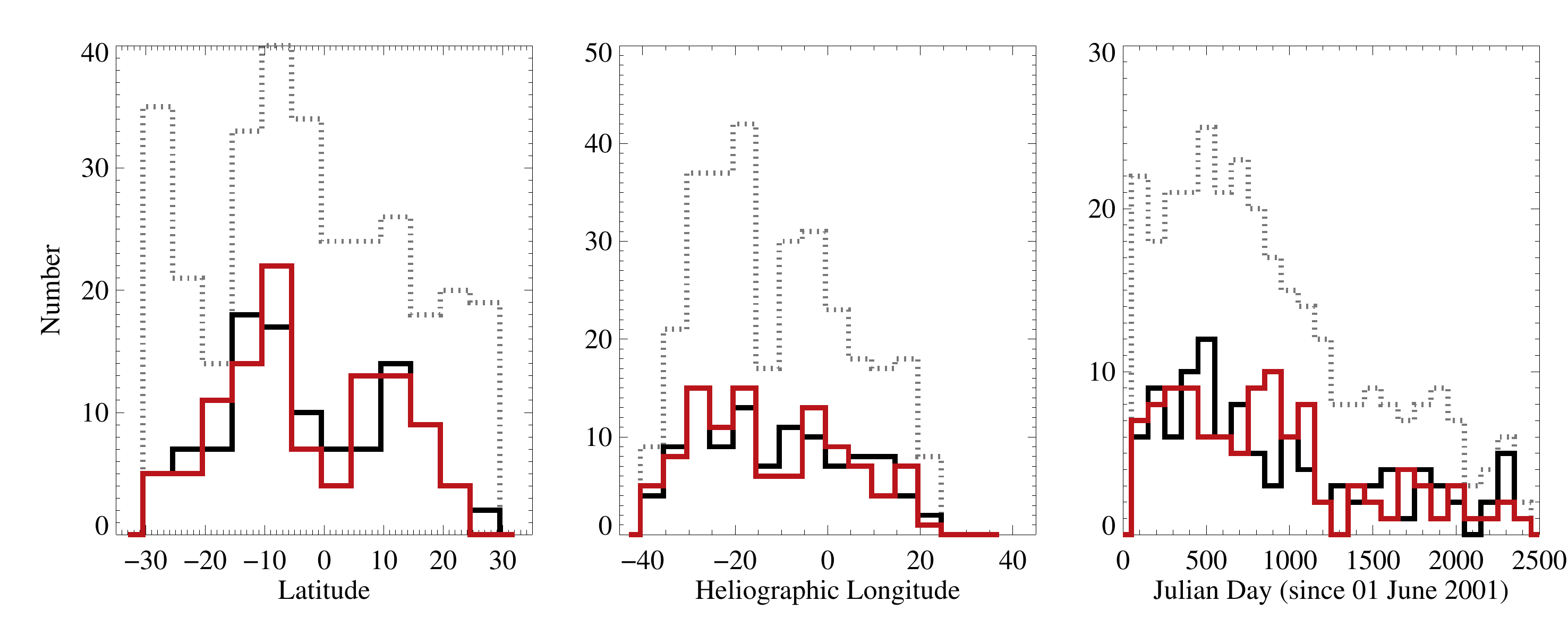}
\plotone{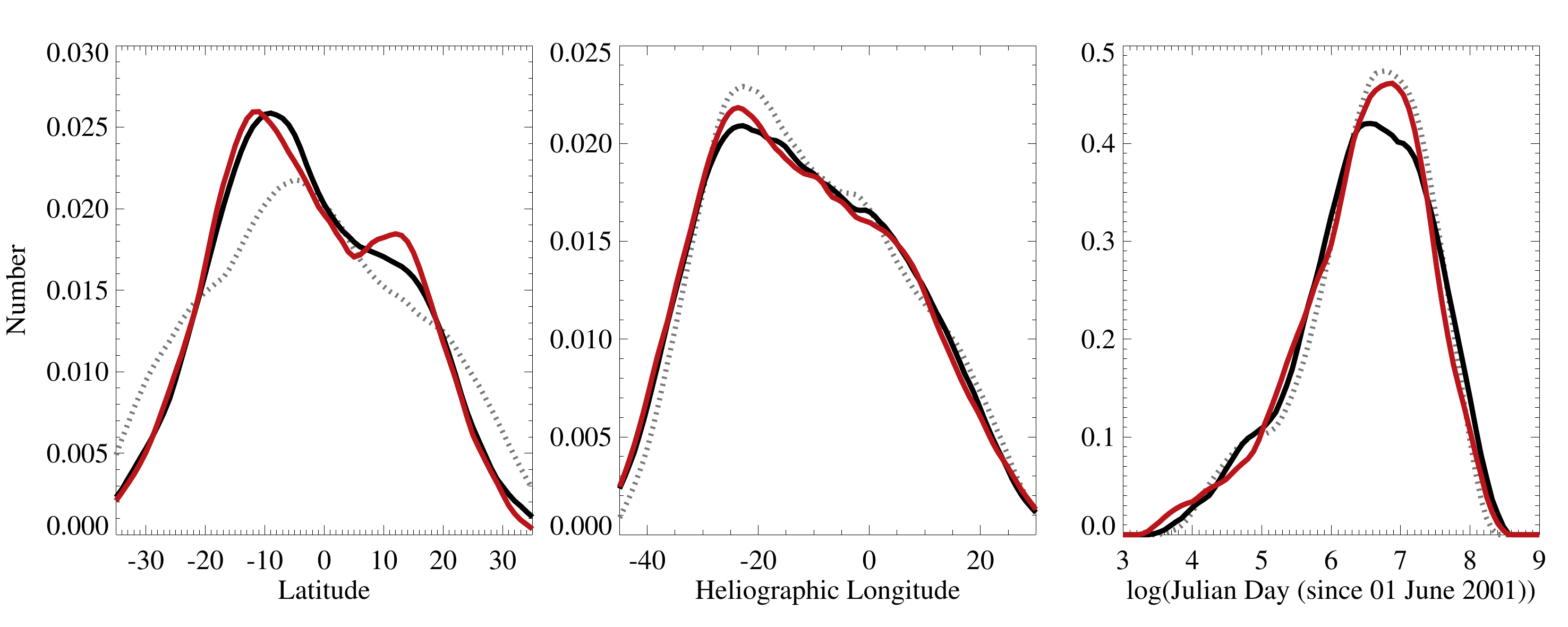}
\caption{Distributions of (left) latitude, (center) heliographic longitude and (right) date
at $t_0$, the defined emergence time.
Shown are the PE distributions (red), the larger sample of NE data (black, dotted) from 
which the matching algorithm drew the final sample (black, solid) which minimized
the integrated difference between the PE and NE Non-Parametric Density Estimates
of the three quantities simultaneously.
Top row: histograms of the relevant quantities, hence indicating number in each
bin; Bottom row: the NPDE distributions, on which the minimization was performed.   The 1-D matches (one variable at a time)
are shown here, whereas the optimization was performed on all three variables simultaneously.
Hence, while better 1-D matches may certainly be obtainable, it would be 
at the cost of the 3-D match results. }
\label{fig:dists}
\end{figure}

An algorithm was developed for post-facto selection from a larger sample of
controls (NE) to match the distribution of the targets (PE) simultaneously in
latitude, longitude, and time.  A non-parametric density estimate \citep[NPDE;
e.g.,][]{Silverman1986}, using the Epanechnikov kernel
and the optimal smoothing parameter for a normal distribution, was used to 
estimate the probability density function for the three
variables on a regular grid in longitude, latitude, and $\ln({\rm time})$
(see Figure~\ref{fig:dists}).
The non-parametric approach was used to
avoid misrepresenting non-Gaussian distributions such as the latitude of
emergence (which is decidedly and expectedly double-peaked);  similarly,
the logarithm of the time variable\footnote{Specifically, the logarithm of the number of
Julian days since 2001 July 25, two days before the first dataset.} was
used to compensate for its extremely skewed distribution.  A simulated
annealing algorithm \citep[e.g.,][]{numrec,Metropolis1953,Kirkpatrick1983}
was employed to select the subset of NE of a specified size (equal
to the number of PE) that minimizes the integrated absolute value of
the difference between the two NPDEs (NE and PE).  Using the integral
preserves the general shapes of the distribution rather than (for example)
employing a peak or maximum difference as a Kolmogorov-Smirnov test
would do.

The results of this matching exercise are shown for the three variables
in Figure~\ref{fig:dists}.  A table listing coordinates for the final
NE targets is provided in Table~\ref{table:NE}, where we list the MDI
orbits generally containing the NE region, the mid-point of the GONG
day used for analysis (see Section~\ref{sec:gongprep}, below), and the
coordinates of that mid-point (note we do not list $t_0$ precisely,
but it is fairly inconsequential).

The equal sample sizes of PE targets and NE controls impose a specific
requirement on the statistical tests: the prior probabilities, of
which type of event (PE or NE) is more or less frequent, is set to
be equal.  This statistical requirement is maintained even after a
further restriction is placed on the data for acceptable GONG duty-cycle
(see Section~\ref{sec:gongprep} and Table~\ref{table:dutycycle}) which
in fact creates small inequities in the sample sizes.  With equal prior
probabilities, the goal of determining whether these populations differ
is emphasized.  Were this a test of prediction, the sample sizes (hence
prior probabilities) should reflect the chances of any random place on
the Sun being a location and time of emergence; clearly this is a ratio
of many thousands to one.

\subsection{Preparing the Doppler Velocity Cubes}
\label{sec:gongprep}

\begin{figure}
\plottwo{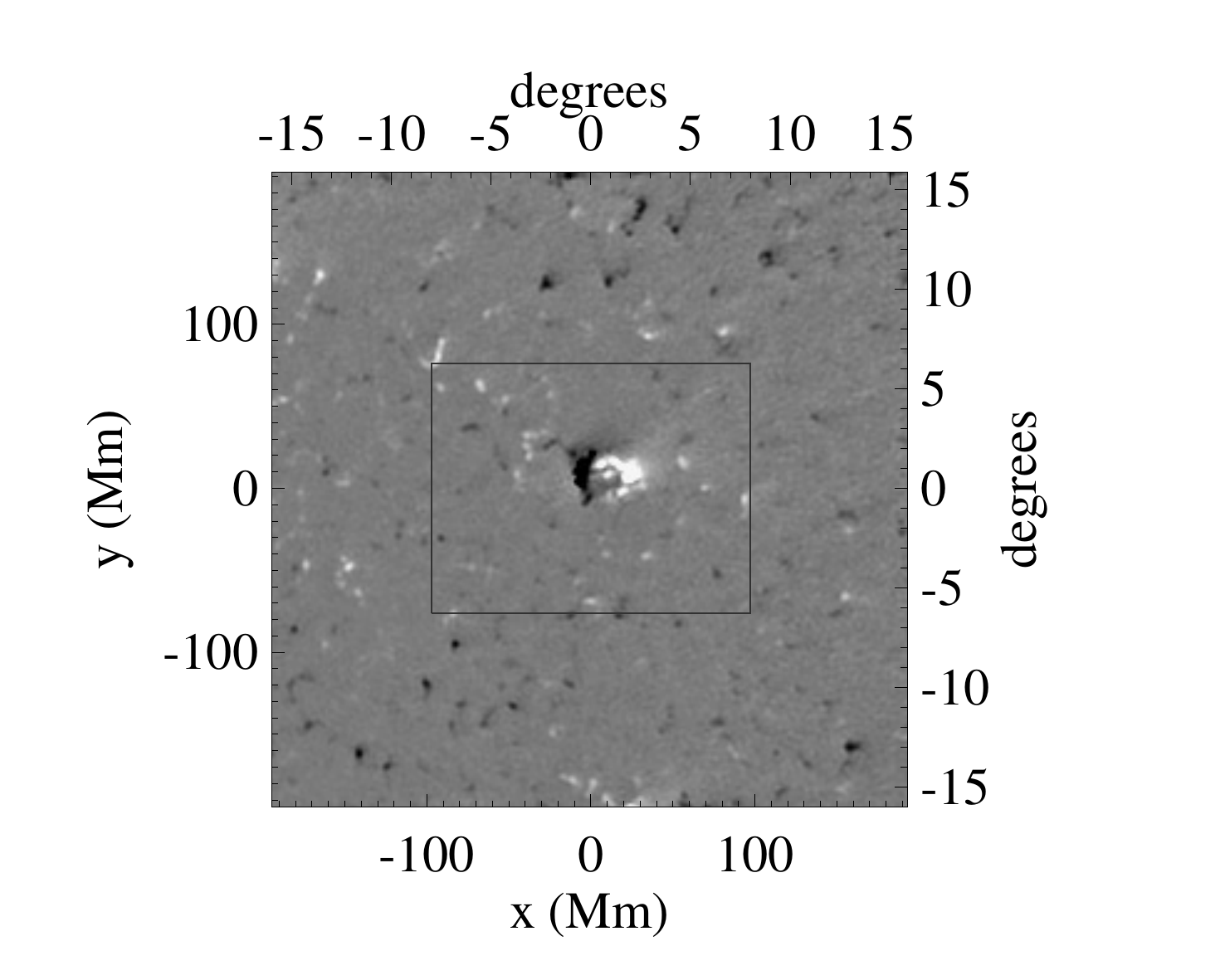}{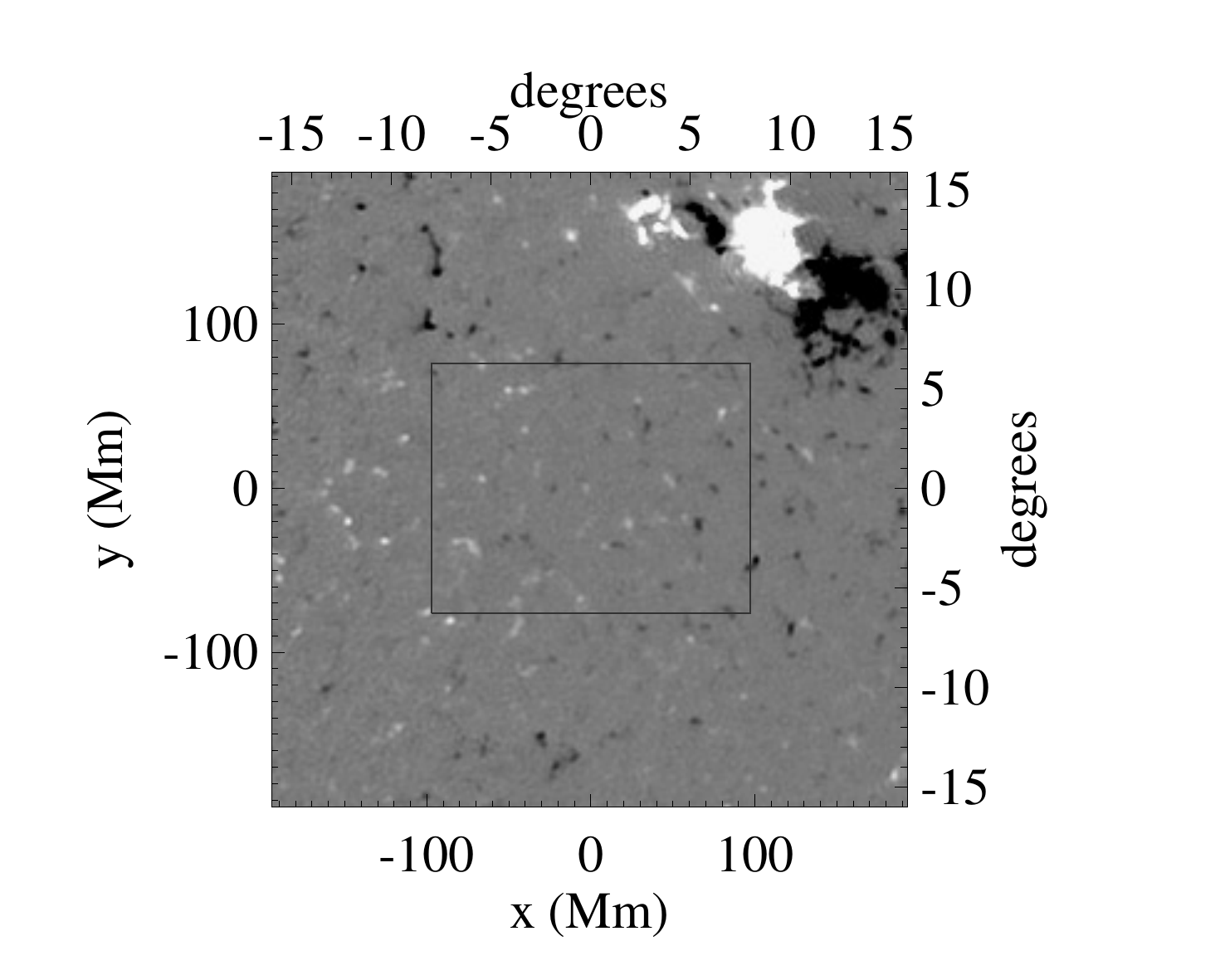}
\plottwo{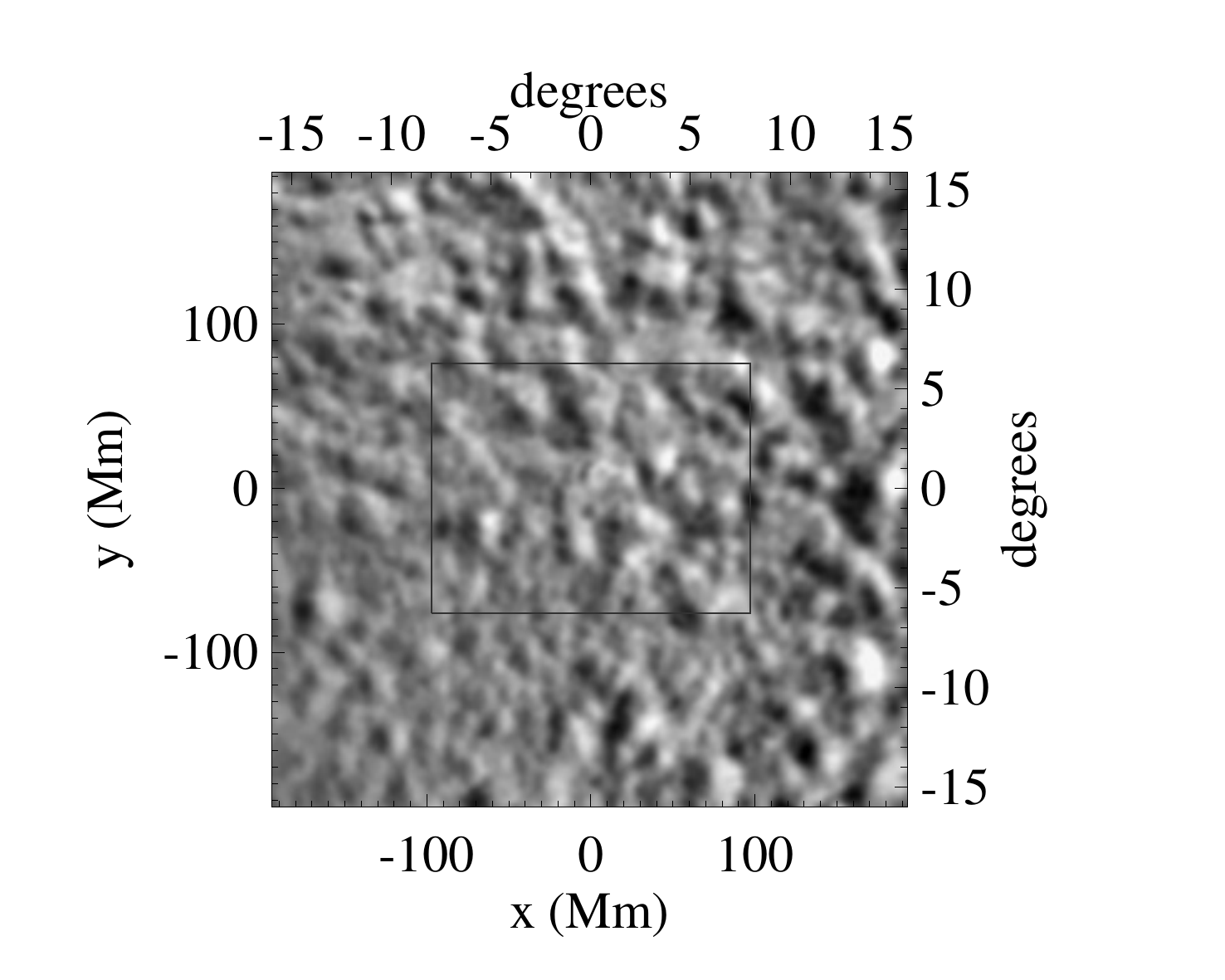}{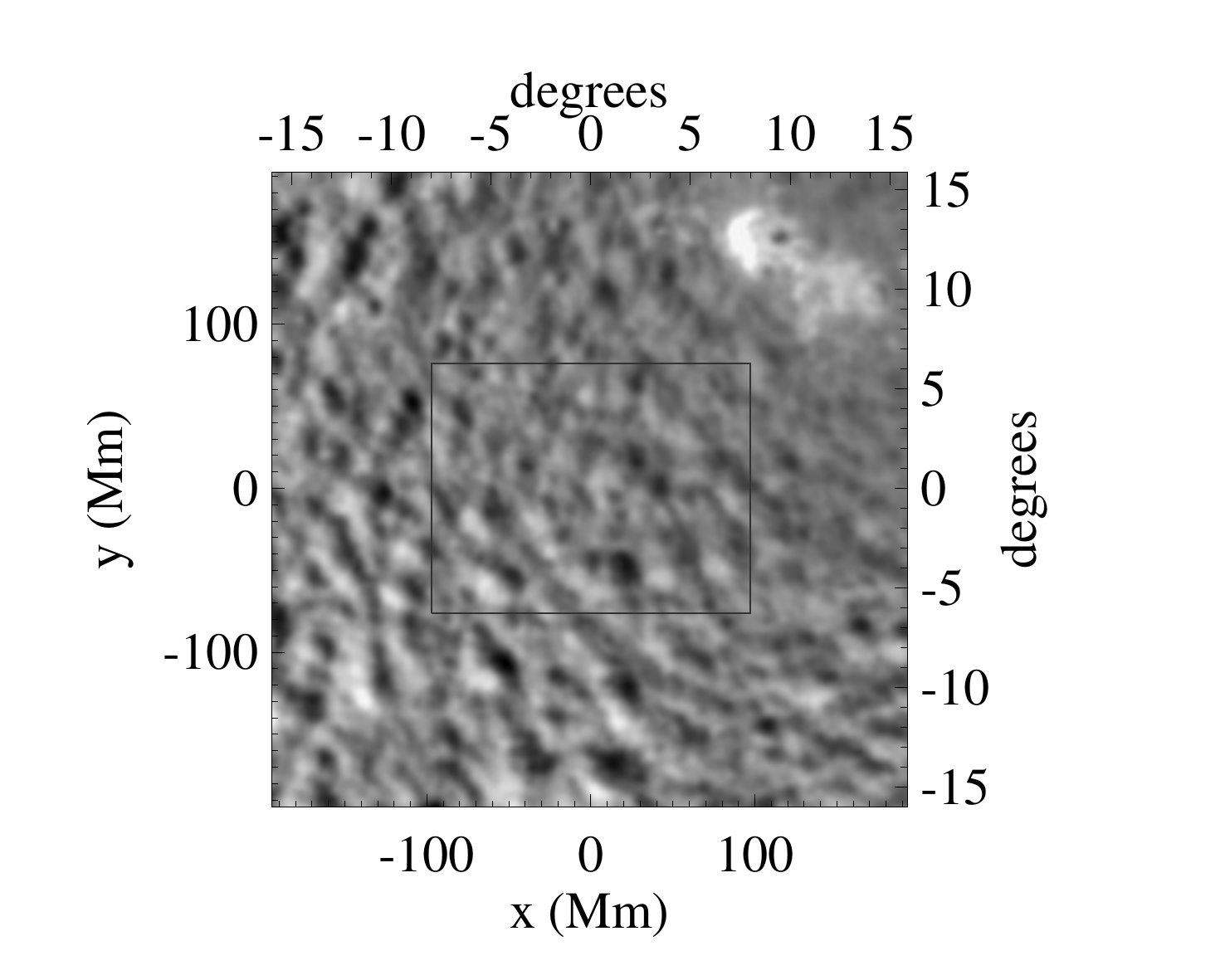}
\caption{Top: The $32^\circ \times 32^\circ$ radial field image, matched to the 
GONG data area, for the same targets as Figures~\ref{fig:t0imagesPE} and~\ref{fig:t0imagesNE}.
Axes are shown in both Mm and degrees from the center tangent point.
Left: average of the PE target AR 10559 2004 February 13 23:59 -- 2004 February 14 06:23,
Right: average NE-target field 2004 February 07 01:35--08:03 UT. 
Bottom: average GONG Doppler images for the same targets, for 
384 minutes each (the length of an interval used for the helioseismology
analysis): Left, for PE target AR 10559, for the same interval as the
magnetogram average above, and Right: for the NE target, and the same interval.
All images: grey boxes indicate the approximate area used for 
initial diagnostics (as in Figures~\ref{fig:t0imagesPE},~\ref{fig:t0imagesNE})
for reference, and as an explanation of the presence of significant magnetic
flux, for example, in many NE targets.
}
\label{fig:32deg}
\end{figure}

After the appropriate target selection, there is no difference in
the treatment of the PE and NE data-cubes produced from the GONG
Doppler velocity data.  Cubes $32^\circ \times 32^\circ$ in extent
were tracked at the Carrington rate, and extracted from
the GONG 1-minute velocity data \citep{gong_pipeline}.  As indicated
in Figures~\ref{fig:cartoon} and~\ref{fig:32deg}, this extracted area
is larger than the original $128\times100$ MDI-pixel area used for
initial evaluation.

The final cubes used for this analysis are one ``GONG-day'' long (1664 min.);
for the PE data, the cubes end 16 minutes after the emergence time $t_0$
due to a small communication error; given the temporal sampling of the magnetic
field data, we do not assign significance to the 16 minutes aside from 
assuming there will be early emergence magnetic flux appearing near the end
of the GONG-day.

\begin{figure}
\plotone{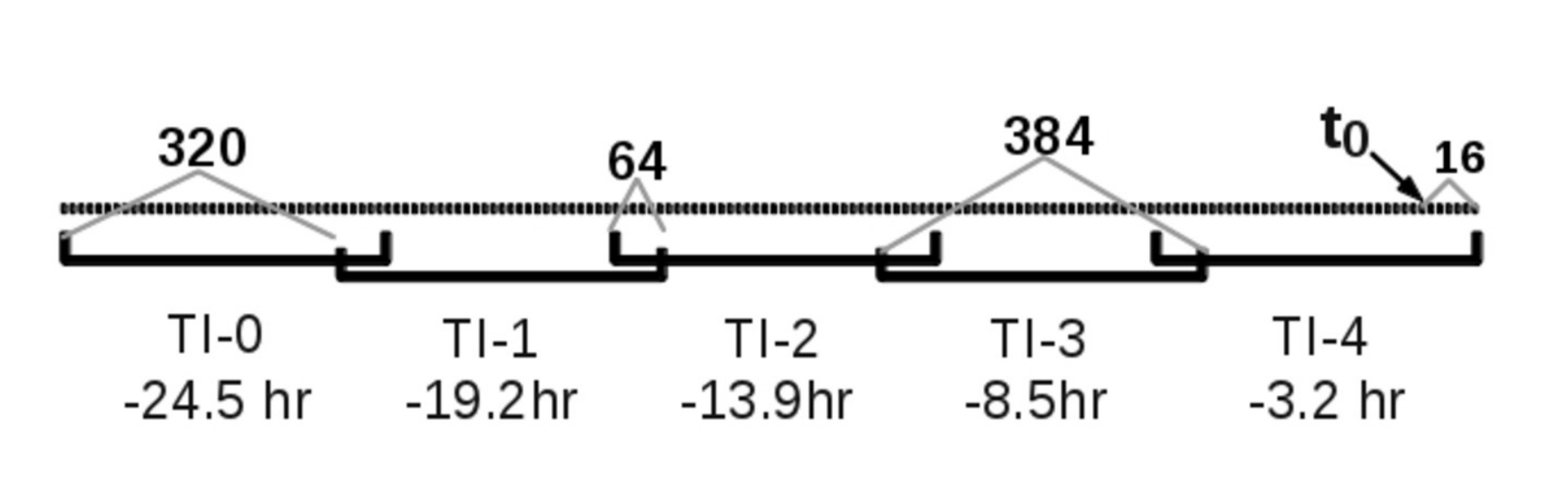}
\caption{A schematic which demonstrates the temporal relationship of the time intervals.
The dotted line represents the time-series of GONG Doppler data, 
bold-face numbers across the top are in minutes; the five time intervals are labeled 
``TI-\#'', and the central
time of each interval, in hours relative to the end of the GONG data,
is indicated below its label.
The GONG data run 1664 minutes, and end 16 minutes after the emergence
time determined as described in the text.  Intervals start every 320 minutes,
are 384 minutes long, and overlap with neighboring intervals by 64 minutes.
This schematic applies to both PE and NE data, albeit with a ``fake'' $t_0$
for the NE targets which corresponds instead to exactly the end of the GONG data.}
\label{fig:timeline}
\end{figure}

The extracted Doppler-velocity data are re-projected using a Postel projection 
\citep{Pearson1990}.
The 1664-minute timeseries are then broken into five time intervals,
each 384~minutes long but starting every 320~minutes (thus an
overlap of 64~minutes between each interval).  A schematic of the
data and the temporal relationship between time intervals is shown in
Figure~\ref{fig:timeline}.

The GONG facility includes different observing sites whose data are
combined to create full temporal coverage.  While the average duty
cycle for GONG data is very high, at times the coverage falters for a
variety of reasons.  Intervals which fall below a duty cycle of 80\% are
not included in the analysis.  This restriction removes data randomly;
there is no reason for duty cycle to be tied to PEs preferentially over
NEs, especially after the matching was performed for location and date.
In addition, what are removed from consideration are individual intervals
rather than an entire PE or NE target.  Table~\ref{table:dutycycle}
presents the resulting sample sizes for PE and NE populations by
interval, after removing data with insufficient duty cycle.

\subsection{The Accompanying Magnetic Data for Analysis}
\label{sec:mdiprep}

In addition to the considering each event (or lack thereof) as viewed by
helioseismology, to confirm that the results are a result of subsurface
processes, we produced a complementary data set of the surface field.
For analysis we attempt to mitigate projection
effects present due to the fact that the MDI data detect only the
line-of-sight component of the flux density (explained in detail below).
We also want to match the measure of the surface magnetic field to the
area and projection used with the GONG Doppler-velocity data cubes.
To achieve this, first the location and $32^\circ \times 32^\circ$
spatial extent of the GONG cubes were identified in MDI data covering
the same time interval.

To minimize projection effects and, more adroitly, use the most
physically meaningful magnetic measure available from the MDI data,
we use a potential-field calculation to retrieve an estimate of the
radial component of the field.  Specifically, the potential field
was calculated to directly match the observed line-of-sight boundary
\citep{Sakurai1982,Bogdan86,Rudenko2001a}, rather than assuming the
boundary was equivalent to the radial component of the field.

In general, the radial component of a potential field (without a source
surface) in the volume above the solar surface can be expressed in a spherical
harmonic expansion as 
\begin{eqnarray}
B_r(&r& \ge R_\odot,\mu,\phi) = \sum_{l=1}^\infty \sum_{m=0}^l (l + 1) 
\bigg ({R_\odot \over r} \bigg )^{l+2}  \\
\times&& \bigg [g_l^m \cos m \phi+ h_l^m \sin m \phi \bigg ] P_l^m(\mu) \; , \nonumber
\end{eqnarray}
where the $P_l^m$ are the associated Legendre functions, $R_\odot$ is the solar
radius, $r$ is distance from the center of the sun, $\theta$ is co-latitude,
and $\phi$ is longitude (measured for any choice of the polar axis). 
Following the approach of \cite{Rudenko2001a} by taking the polar axis of the
coordinate system to lie along the line of sight, and using relationships among
the associated Legendre functions as done by \cite{Bogdan86}, the coefficients
can be written as
\begin{eqnarray}
g_l^m &=& {(2 l + 3) (l - m)! \over 4 \pi (l + m + 1)!} \int_0^1 d\mu\\
 && \times\int_0^{2\pi} d\phi \, \cos m \phi P_{l+1}^m(\mu) B_l(R_\odot,\mu,\phi) \nonumber
\end{eqnarray}
and 
\begin{eqnarray}
h_l^m &=& {(2 l + 3) (l - m)! \over 4 \pi (l + m + 1)!} \int_0^1 d\mu \\ 
&& \times \int_0^{2\pi} d\phi \, \sin m \phi P_{l+1}^m(\mu) B_l(R_\odot,\mu,\phi) \; , \nonumber
\end{eqnarray}
where $B_l(R_\odot,\mu,\phi)$ is the line of sight component of the field
at the solar surface. To ensure that the monopole term vanishes in the
sum, we further assumed that the field on the far side of the Sun was
given by $B_l(R_\odot,\pi-\theta,\phi)=B_l(R_\odot,\theta,\phi)$,
where the front side of the Sun is assumed to lie in the range
$0<\theta<\pi/2$. This can produce some unphysical results
very close to the limb, but does not greatly affect the field
at the surface in the restricted area of the disk considered in
this investigation.  The integrals were evaluated using a simple
trapezoid method, and the spherical harmonics were computed using
the freely available software archive SHTOOLS\footnote{available at
{\tt http://www.ipgp.fr/$\sim$wieczor/SHTOOLS/SHTOOLS.html}}, which
have a relative error of less than $10^{-5}$ up to degrees of at least
2600. However, only terms up to degree of 1000 were included, as this is
sufficient to reconstruct spatial scales on the order of the resolution
of MDI.  Note also that the acoustic modes in the GONG data are seen up
to about $l=1000$ \citep[see Figure~1 from][]{trt_p2}. 

The above calculations were performed on an extracted cube slightly larger
than $32^\circ \times 32^\circ$, then the potential field radial component
was subjected to Postel-projection and trimmed to exactly match the GONG
datacubes.  Hence, we have for each PE and NE data set, a time-series
of the radial component of the field matched in area, and matched in
projection, to sub-surface observations made by helioseismology, albeit
the latter by a different instrument.

From these maps, an appropriate time-series of the history of
the field at the target and its immediate surroundings is computed,
for comparison with the results of helioseismology.  Sample pairs
of average radial field density and average corresponding Doppler 
data are shown in Figure~\ref{fig:32deg} for the PE and NE
examples of Figures~\ref{fig:t0imagesPE},~\ref{fig:t0imagesNE}.

\begin{figure}
\plotone{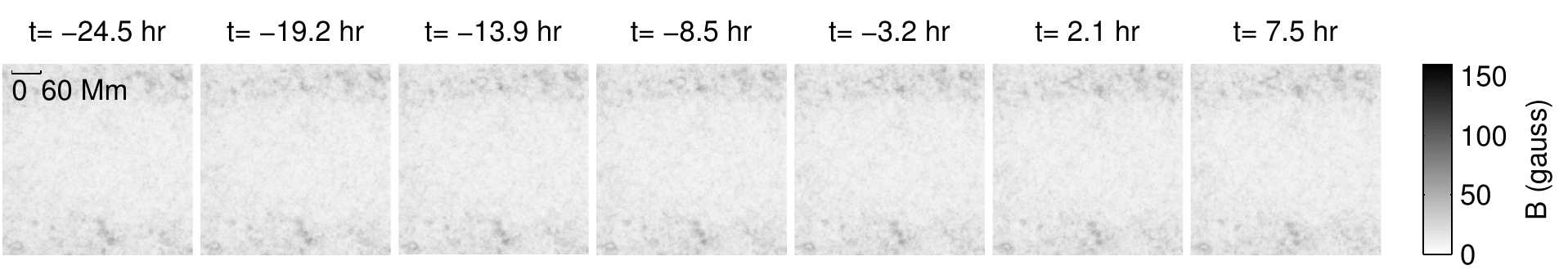}
\plotone{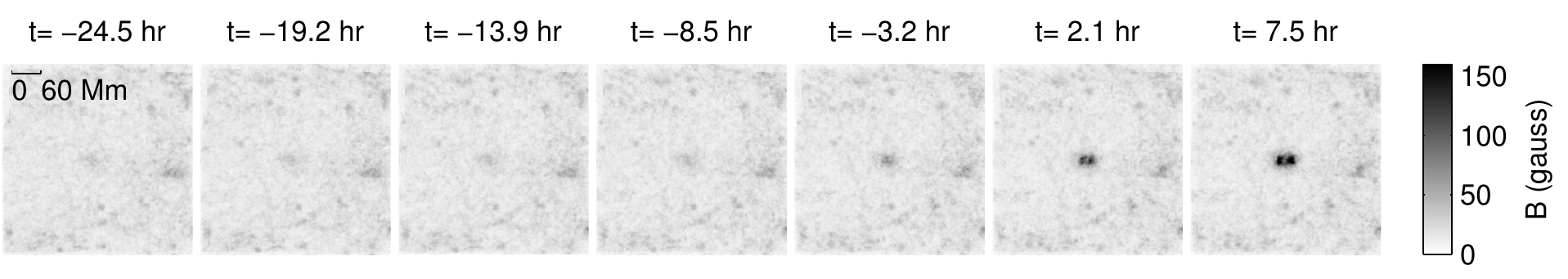}
\plotone{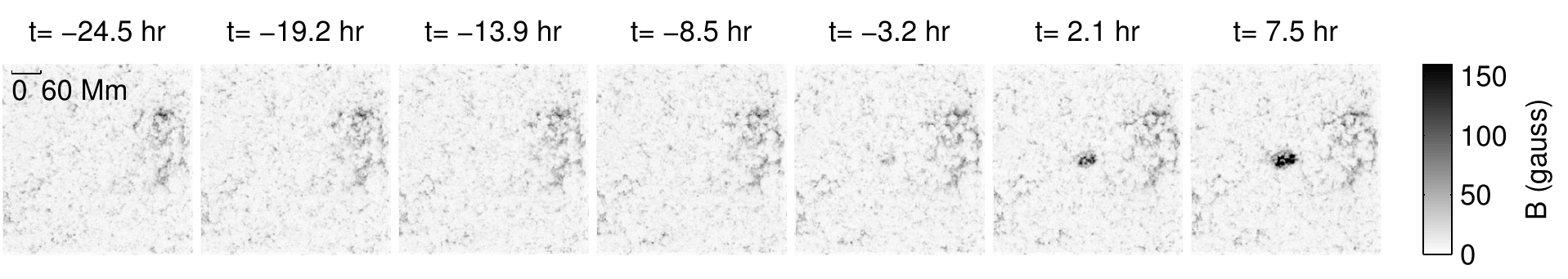}
\caption{Averages over all samples of the unsigned radial field for each of the time intervals
as accompanies the seismology data.  In addition to the five primary 
intervals prior to emergence, we show here the averages for two additional
intervals post-emergence, for comparison.  Times indicate the central
time of each interval, following Figure~\ref{fig:timeline}.  All figures use the same grey-scale.
Top: the NE samples, Middle: the PE samples, Bottom: the ``Ultra-Clean'' subset
of PE samples.}
\label{fig:Baverages}
\end{figure}

For these accompanying magnetic data, we show in Figure~\ref{fig:Baverages}
the unsigned radial field averaged over all samples, for each of the time intervals
used for the seismology analysis.  To provide context, we extend this slightly in time
and show the averages for two additional post-emergence time intervals.
Of note are the distinct lack of variation in the NE data, but also 
the noticeable bands of stronger signal at the top and bottom of the 
NE data cubes compared to the central portion.  The PE data show a distinct
early signature of surface field 24hr prior to the emergence time, and
a clear bipolar signature after emergence is underway.  The biolar 
structure is less clear but arguably present in the subset of PE
data, the early signature is arguably completely absent when only the cleanest, 
``most virgin'' examples were chosen.  At the same time, averaging over a smaller
number for the ``ultra-clean'' dataset allows a single sample to influence the average: 
the strong persistent signal on the right-hand portion of the ``ultra-clean'' mean in 
Figure~\ref{fig:Baverages} is primarily due to a strong plage area near NOAA~AR9645.

\subsection{Further Corrections}
\label{section:centroids}

The latitude reported by NOAA was generally unchanged for extracting the GONG
data cubes; the longitude was obviously updated according to $t_0$.  
For the later analysis, especially the averages over all samples used
in \citet{trt_p2}, the coordinates were refined in the following manner.
The time-series of the radial magnetic field were used to construct
bitmaps of new flux using the difference ($|\delta B|$) between the field
roughly 12 hr after $t_0$ and the first time interval (roughly 24 hr before $t_0$),
and only including in the bitmap areas where $|\delta B| > 0.3 \times {\rm max}(|\delta B|$).
A centroid was created from the resulting bitmap, and the coordinates
were then assigned to be the location of this centroid.

In this manner, the analysis which is performed on averages taken over
space, time, or sample, will provide results that are not diluted by 
subtle differences in emergence location within the field of view.
Accordingly, no similar refinement was performed for the NE samples, as
there are no events by which to define such a refinement.

\section{Statistical Contamination Issues}
\label{sec:contamination}

This is a study trying to detect a small difference between two populations.
How these populations are defined and the samples
obtained, then, will directly affect the reliability of the results.
The goal is that the PE regions be clear, distinct, isolated,
fairly near disk-center emergence episodes, and the NE regions be
emergence-free episodes matched to the PE distributions in location
and time (effectively, solar-cycle activity level) as described above.

Statistical contamination, the existence of a bias that will inadvertently
identify the two populations without being directly related to the 
emergence process, may take a variety
of forms.  Alluded to in section \S~\ref{sec:design}, we describe below
our understanding of various contributions to possible contamination,
and our efforts to mitigate them.

\subsection{Nearby Field}

Ideally, the background, nearby, or pre-existing field in the NE
targets (their distribution in space, flux density, total flux, {\it
etc.}) are indistinguishable from that of the PE targets prior
to emergence.  The emergence episodes and non-emergence regions were
initially characterized by $128\times 100$-pixel tracked boxes in the MDI
image-plane coordinate frame.  The data cubes used for analysis were,
as described above, $32^\circ \times 32^\circ$ on a heliographic grid.
The difference between these two can be seen in Figure~\ref{fig:32deg},
and is not insignificant.  The most noticeable effect is that the NE
cubes in fact often contain stronger field at the periphery than the
cut-off used to select the smaller areas (Figure~\ref{fig:Baverages}).
The PE cubes were isolated from nearby active regions in the original
$128\times100$-pixel evaluation, but strong field (active regions)
can be found in the larger $32^\circ \times 32^\circ$ field of view.

\begin{figure}
\plotone{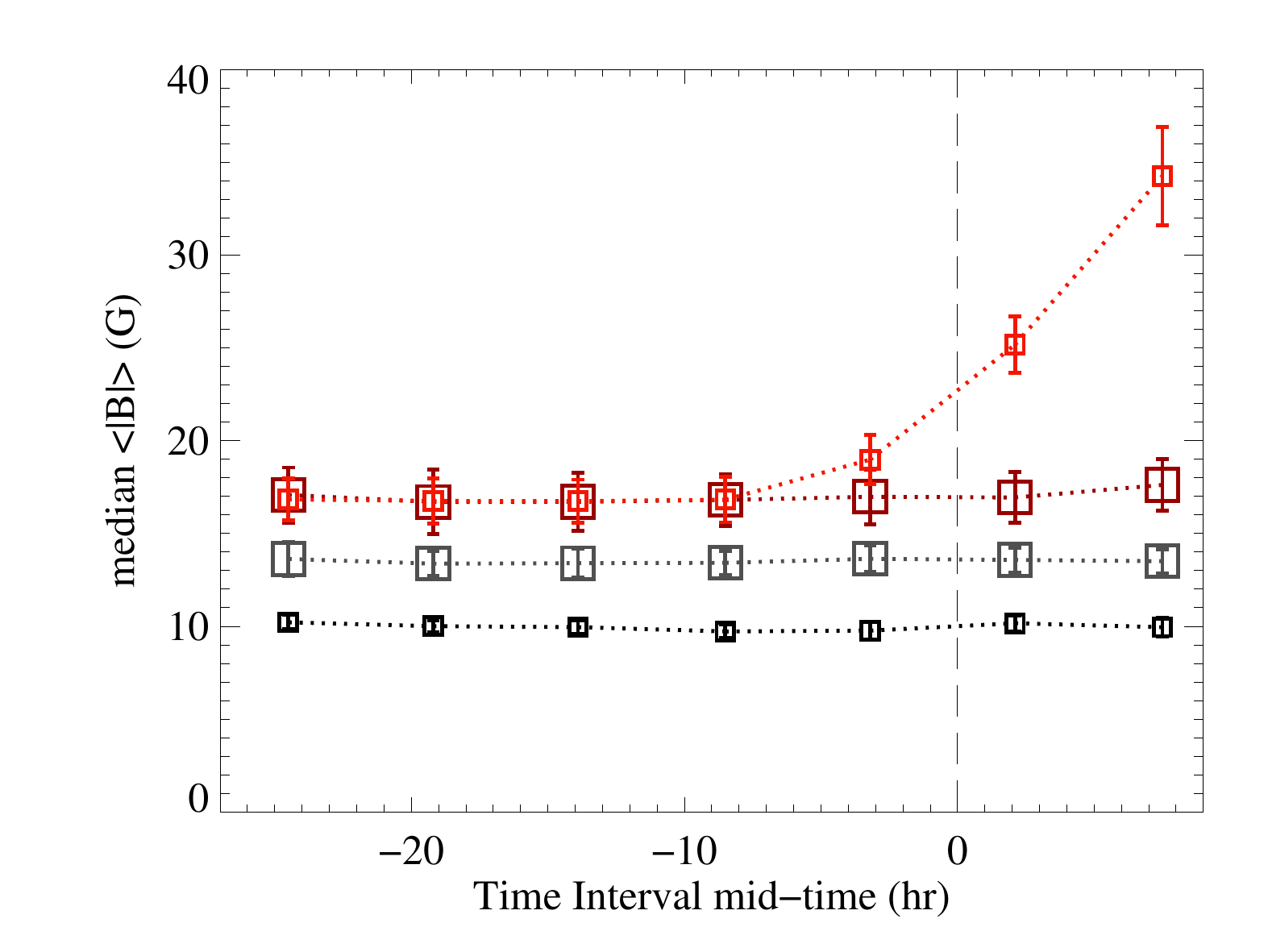} 
\caption{Median of the area-averaged unsigned field and the errors in the 
median (using a bootstrap method), for 
both the PE (red), and NE (black) data, plotted as a function of the 
central time of the intervals relative to the end of the GONG day (which is 
effectively $t_0$).  Larger symbols (and muted red/grey) indicate
that the median was taken over the entire extracted area, smaller (red/black)
symbols indicate that only the smaller $\approx16^\circ \times 16^\circ$ 
area used for the helioseismology analysis was included.}
\label{fig:averageB}
\end{figure}

By comparing the signals averaged as shown in Figure~\ref{fig:averageB},
it is clear that there is a bias: the median signal of magnetic field
is larger in the PE samples as compared to the NE samples.  Note that by
showing the median, rather than the mean, the results are not influenced
by outliers and the distributions display a real difference.  For the full
$32^\circ \times 32^\circ$ field of view, the difference is significant
but not large; when considering only the smaller central $16^\circ
\times 16^\circ$, the PE sample result does not change noticeably
(until emergence begins in the last time interval), whereas the NE
sample median signal is quite reduced.  This confirms that the initial
$128\times100$-pixel evaluation area for the NE sample is ``too quiet''
compared to the enhanced signal in the NE sample peripheries and to the
PE sample, even though the selection threshold was a generous 1kG
(Section~\ref{section:quiets}).

The source of this bias may be introduced or it may be a real effect.
The emergence really could start more than a day before $t_0$, in which
case there is no error, just a real physical effect only visible in the
ensemble.  However, by imposing a field strength limit on the NEs but
not PEs, we may have introduced an artificial bias into the samples.
Due to the matching in latitude and longitude, there should be no
gross preferential prevalence of ``background'' field as there would
be had all of the NE regions, for example, been selected outside the
active latitudes or all in the same hemisphere.   However, there was
also no de-selection of PE candidates based on ``active longitudes''
\citep{PetrovayAbuzeid1991,gaizdeltaI,PojogaCudnik2002}, and active
longitude lifetimes are likely too short to be captured simultaneously in
the time-matching and longitude-matching.  If the bias is the effect of
active longitudes, this is a real (solar) bias towards having pre-existing
field for the PEs.  The fact that the NEs are ``too quiet'' implies that
the inconsistent use of a threshold contributes to the bias, but may
not be the only effect.  The significance of this systematic difference
between the samples is discussed in detail in \citetalias{trt_p3}.

\subsection{Nearby or Short-Lived Emergence}
\label{sec:er}

It is conceivable that nearby or on-going short-lived
flux emergence may contaminate the seismology signal we search for.

No screening or diagnostics were performed to specifically rule out
nearby emergence episodes for either PE or NE samples.
Both may have emerging flux regions in the periphery, and these datasets
were not removed from consideration (as long as there were no emerging
flux regions within the central $16^\circ \times 16^\circ$, or $\approx
100\,{\rm Mm} \times 100\,{\rm Mm}$).  A variety of seismic analyses will be
performed with varying pupil sizes \citep{trt_p2}, thus the influence of field
in the sample peripheries can in fact be studied.  

Very small-scale short-lived emergence episodes, ``ephemeral regions''
are ubiquitous and bring substantial flux to the solar surface
\citep{HarveyZwaan1993,hagenaar_etal_2003}.  The presence of ephemeral regions
is not selected for or against, as their peak field strengths generally fall
below the NE-selection threshold of 1kG in MDI data, except one or two cases of
removing an NE candidate due to a long-lived or especially large ephemeral
region occurring at the center of the target.  We make the assumption that the
rate and distribution of ephemeral regions is the same between the samples of
the NE and PE populations, and propose that no statistical bias is
introduced due to the presence of ephemeral regions.

\subsection{Mis-Determination of Emergence Time}
\label{sec:mess}

Numerous sources of error could lead to a mis-determination of
$t_0$, with effects presenting as bias or as random error.

The coarse temporal resolution MDI data used here could lead to a
significant amount of ``new'' surface flux being present for an hour or
so before the ``emergence time'' $t_0$.  The limited spatial resolution
of the MDI data could lead to a significant amount of undetectable flux
being present for an unknown period before the ``emergence time'' $t_0$.
``Significant'' is used here qualitatively, because it is the lack of data
which is the primary source of the uncertainty itself.  Lack of adequate
sampling should add an element of random noise to comparisons between
segments.  The reliance on line-of-sight data, however, may present
a systemmatic late determination of $t_0$ with respect to observing
angle, since early flux emergence is signaled by horizontal field
\citep{Zwaan1985,ZhangSong92,twist,Bernasconi_etal_02,kubo_etal_2003}.
In and of itself, the instrumental limitations should not present a
statistical contamination between the NE and PE samples.  Since the
presence of surface field when none is expected (as due to the
mis-determination of $t_0$) may impact the Doppler signal and hence
the inferred helioseismic parameters, the results for the time interval
comprising the last hours prior to $t_0$ will be interpreted with this
uncertainty taken into account \citep{trt_p2,trt_p3}.

Of a more subtle nature, in terms of this study, is the nature of
flux emergence itself, the early evolution of active regions, and
whether or how a very young active region is distinguishable from the
general evolving magnetic background.  While we employed an objective and
quantitative method to determine $t_0$, as needed for a statistical study,
upon examination of any individual case, $t_0$ could be argued with.
An example is shown in Figure~\ref{fig:t0imagesPEmess}.  An area of
unchanging plage is co-spatial with the eventual emergence of NOAA AR
9564, and episodes of small bipoles appearing are evident prior to $t_0$
upon detailed inspection.  These bipoles would not gain attention beyond
the numerous ephemeral regions continuously appearing on the surface
(see Section~\ref{sec:er}) and indeed they did not gain NOAA's attention,
except that they were located where NOAA AR 9564 eventually appeared.

We hypothesize without further investigation, that the
pre-emergence surface field signature in the all-PE averages
(Figure~\ref{fig:Baverages}) is an indication of this very common
characteristic: pre-emergence field can be present, whether as
remnant plage or very early emergence episodes that are un-notable
in any individual PE time series.  As commented on earlier, when
only examples are selected for which -- by visual inspection --
there is no pre-emergence surface field, the pre-emergence field
signature is reduced if not absent.  A third option is that very early
emerging flux is distributed and weak, and detectable only on average
(Figure~\ref{fig:Baverages}) with the MDI data due to the significantly
reduced noise; in this case the pre-emergent surface field signature
is absent for the clean subset not due to their ``ultra-clean'' nature,
but due to the smaller number of datasets being averaged, and hence the
increased noise (compared to averages for all regions).

The impact of a varied flux-emergence rate on later analysis should
be a source of noise but not statistical contamination.  The rate of
emergence of new flux was cited in \citet{ilonidis_etal_2011} as a key
parameter relating to the strength and timing of the pre-emergence signal.
However, it does not bias the NE {\it vs.} PE samples.

The final evaluation is that $t_0$ may be mis-determined by an amount
comparable to the MDI 96-min sampling, hence the final time interval
used for helioseismology analysis will be assumed contaminated
by early emergence.  Smaller episodes of new flux appearance are
indistinguishable from that which routinely occurs over the solar
disk without the subsequent formation of an active region, and can simply
be considered a source of noise for the present analysis.

\section{Discussion}
\label{sec:discussion}

The tools of local helioseismology rightfully hold hope of sensitive
and powerful diagnostic tools of the solar sub-surface structure,
evolution, and behavior.  To interpret the helioseismic signals with
physical insight, they must be isolated to those relevant to the events
in question.  To fruitfully make use of the signals, the full extent
of bias and contamination must be understood.  We have designed a study to
examine what signatures prior to the appearance of solar active regions
may be detected by local helioseismology tools and data at this time,
and outlined the data selection criteria and preparation herein.

This study focuses on determining whether or not a seismology signal is
evident prior to emergence and what its character might be.  The goal
is, as discussed in Section~\ref{sec:intro}, inferring changes in the
subsurface associated with active-region formation.   Based on the
preparation described here, in \citet{trt_p2} we report on average
sub-surface properties of the two samples (PE and NE) as derived using
helioseismic holography, and find statistically significant signatures
in average subsurface flows and wave speeds, but do not detect evidence
of strong spatially extended flows in the top 20~Mm during the day
preceding visible emergence.  In \citet{trt_p3}, parameters are derived
from the seismology and magnetic field to characterize each of the PE
and NE regions, and discriminant analysis is used to measure differences
between the sample sets.  While statistically significant differences are
found from this analysis, it is found that no single parameter can clearly
distinguish a pre-emergence from a non-emergence for any single region.

To mitigate sources of bias, the distributions of the samples are matched
in location and time (epoch within the solar cycle).  This approach is
novel; however Pre-Emergence areas are targeted here exactly because they
{\it did} form an active region significant enough to be noticed by NOAA.
The PE targets can thus be studied with respect to the {\it known}
location and time of emergence, and the magnetic- and seismology-based
analysis performed with respect to the target's {\it known} coordinates.
There is a random component in the selection of the No-Emergence regions,
but they, too, are selected with knowledge that no emergence occurred
within a specific time interval.  Hence, there is a bias in that we
are pre-selecting targets for study according to what is known to have
happened.

There is an intrinsic difference between this study design and any attempt
at ``forecasting'' the emergence of an active region.  A forecasting
study would instead be required to sample all possible emergence sites
and compare the signals to all other possible sites, without {\it a
priori} knowledge aiding the analysis methods.  At the very least, a study
designed for forecasting must employ samples and statistics which reflect
the prior probability of an active region emerging at a randomly selected
place and time over the observable disk, which is extremely small.
While the results presented in \citet{trt_p2} and \citet{trt_p3}, and the
available ``blind'' datasets (see Section~\ref{sec:availability}, below)
may serve to guide later studies of the true forecasting ability of
seismology for active region appearance, we caution that study design
and attention to prior probabilities are crucial to answering specific
questions posed.

As a study with a fairly large sample, comparisons between this
(and \citet{trt_p2} and \citet{trt_p3})
may be made to e.g., \citet{komm_etal_2009,komm_etal_2011};
however, the definition of ``emerging'' differs, in that here we
focus on the period prior to any surface field --  whereas the
earlier studies included both new and growing active regions (with
surface flux present).  As such, the present study may be seen as
an extension of case-studies which also focused on pre-emergence
periods \citep[e.g.,][]{jensen_etal_2001,ilonidis_etal_2011,Braun2012}
to statistically-significant sample sizes, however the methods and
interpretive tools (depths, cadence, control samples if any) differ
between these studies and the present one.  We describe here the
steps taken to acquire both the statistically significant sample
size with a clear focus on pre-emergence phenomena (although see
Section~\ref{sec:future}, below).  With better tools and analysis
approaches, the sometimes conflicting results in the literature should
give way; then, only those effects which are truly specific to the
emergence process will be the focus of discussion.

Any seismic changes detected prior to surface changes will be evaluated
in the context of the predictions made by different theories covering
the source and formation mechanisms of solar active regions.  But the
seismology is influenced by the early surface behavior, the interpretation
of the surface behavior is influenced by our understanding of the
emerging-flux scenarios, which is what we are trying to learn about
using seismology.  The analysis has a circularity to it which implies
one thing most strongly: interpretation must be done with utmost care.
Only then can model predictions be validated.

Emergence scenarios differ between active
regions with respect to rate of flux increase, the existence of distinct
emergence episodes, location with respect to remnant field, {\it etc.}
As mentioned above, the early evolution of active regions is an active
research area and distinctly tied to the sub-surface behavior which is
the focus of this study.

There are efforts underway \citep{SDO_ComputerVision}
to perform automatic feature recognition on data from, for example, the instruments
of SDO. Combining emergence indications from HMI and AIA may be advantageous.
Using such database of emergence times defined by an independent 
algorithm may lend objectivity to the results and ease of acquiring the 
larger samples we suggest, but it must be accompanied by research
on the early evolution of active regions.

\subsection{For Future Studies}
\label{sec:future}

Hindsight enables future improvement.  The flaws of a study design become
distinctly clear as the study progresses and ``issues'' arise; in the best
situations the flaws can be remedied, but in many cases due to resource
limitations, corrections or accommodations must be made mid-course.
Specific effects that the flaws in the present study's design had on
the results will be discussed in \cite{trt_p2,trt_p3} as appropriate.
Whereas this paper discusses the details of the design, the results 
also comprise the lessons learned over the duration of this study:
\begin{enumerate}
\item Characterizing early active-region appearance and evolution
is very much a research topic.  The (objective, independent) determination
of emergence time and location should be performed using, ideally,
vector magnetic field data, to detect the earliest horizontal field
(Section~\ref{sec:mess});
vector data may thus alleviate any systemmatic bias in $t_0$ as a function of observing angle.
Resolution issues aside, the early evolution of active regions may form a spectrum
of behavior such that assigning a single location and time is, in fact, inappropriate.
However, for a statistical study, the determination of emergence time must be
performed, as we did here, in an objective and repeatable manner --
recognizing that the answer is very sensitive to data resolution,
sensitivity, and cadence. 

\item Data selection rules must be applied to areas used in the final analysis
with minimal discrepancies.  As described in Section~\ref{sec:data}, the initial
evaluation of the PE and NE regions was performed on a
much smaller field of view than was eventually extracted from the GONG data (and
than was also eventually extracted from the MDI data for magnetic-field comparisons).
As such, there was more, and more varied, peripheral activity than
was expected in both the NEs and PE samples.  The contamination is inevitable
given the large number of active regions during solar maximum activity; still,
the bands of significant field in the periphery of the NE average magnetograms 
(Figure~\ref{fig:Baverages}) were unexpected.
\item  The distribution of background field must also be matched between
populations in a manner analogous to matching the distributions of 
location and time (epoch within a solar cycle).
That is, there is in fact a bias in the data sets
used here, since the NE regions are overall quieter, with
less background field, than the PEs (see Figure~\ref{fig:averageB}).
Rather than just select for ``no field above a certain threshold''
for the NE regions, areas should be selected which match the pre-emergence
background field distribution characteristics of the PEs.  This task is not
trivial.  
\item Related, the spatial distribution of the field may be important, since the 
seismology signatures are derived from Doppler signals both at the focal
point and in an annulus, whose size relates to the depth sampled.  Regions
emerging into an existing plage area will have a different surrounding
flux distribution than very-quiet non-emergence areas.  Contrariwise, 
if stable plage areas are chosen preferentially as the non-emergence 
targets, then a bias is clearly introduced.
\item Utilize helioseismic data and magnetic data from the same source, if at all 
possible.  It was unfortunate that magnetic field data were not readily available 
for this study from the GONG system.  HMI is the logical data source for 
any follow-on statistical study to what is presented here.
\item Examine 48\,hr or more prior to emergence rather than only 24\,hr (and, of course,
match this for the control data).  This will decrease the number of regions
available within suitable observing angles, however will allow additional
evolution to be detected.
\item For studies that employ statistical analysis, initial target sample sizes
should be 5--10 times larger than assumed sufficient for the final analysis.
The robustness of results depends on noise in the data and the many
sources of bias.  But it also involves an interplay between sample sizes
{\it vs.} the number of variables tested.  The larger the sample size,
and the larger that size is relative to the number of variables under
consideration, the smaller the chance of statistical flukes in outcome.
The initial ``PE'' target list for this study numbered almost 500 regions;
after removing targets due to data problems, significant spatial/temporal
overlap, matching for latitude/longitude/epoch, and accounting for duty-cycle
limitations, each time-interval used had $\approx85-90$ samples.
\end{enumerate}

\subsection{Data Availability}
\label{sec:availability}

Despite the shortcomings identified above, the present study provides a rich
data set for investigating questions of pre-emergence signatures 
of solar active regions, the sensitivity of results to methodology, {\it etc.} 

To that end, we make the datasets used in this study available through
\url{http://www.cora.nwra.com/LWSPredictEmergence \\ /Site/Data\_Sets.html}
(following the link which cites this paper).  
They have been prepared for double-blind tests,
in that the data from both PE and NE samples are available but have
been randomized with all identifying information removed from filenames
and file headers.  Also included at that website will be an uploadable form by which 
to submit ``answers'' to the same Discriminant
Analysis code used in \citet{trt_p3}, so that groups interested in 
direct method comparisons can quantitatively compare performance against 
our published results.

\acknowledgments

The authors acknowledge support from NASA contract NNH07CD25C and the hospitality and support of the National Solar Observatory, and also thank the referee for careful reading and comments. This work utilizes data obtained by the Global Oscillation Network Group (GONG) program, managed by the National Solar Observatory, which is operated by the Association of Universities for Research in Astronomy (AURA), Inc. under a cooperative agreement with the National Science Foundation. The data were acquired by instruments operated by the Big Bear Solar Observatory, High Altitude Observa- tory, Learmonth Solar Observatory, Udaipur Solar Observatory, Instituto de
Astrof{\'i}sica de Canarias, and Cerro Tololo InterAmerican Observatory. MDI data were provided by the SOHO/ MDI consortium. SOHO is a project of international cooperation between ESA and NASA. A.C.B. acknowledges collaborative work within the framework of the DFG CRC 963 ``Astrophysical Flow Instabilities and Turbulence."

\bibliographystyle{apj}
\bibliography{biblio}



\clearpage

\clearpage
\LongTables
\begin{deluxetable}{rrrrr}  
\tablecolumns{5}
\tablewidth{0pc}
\tablecaption{Identifying Coordinates for Pre-Emergence Targets.} 
\tablehead{ 
\colhead{Region} & \colhead{Emergence} & \multicolumn{2}{c}{Location ($^\circ$)} & \colhead{Max Size } \\
\colhead{ID} & \colhead{Date ``$t_0$''} & \colhead{Lat } & \colhead{Long } & \colhead{$(\mu {\rm H})$} 
 } 
\startdata 
9559 & 2001-07-27 04:51 & -26.0 & -30.0 & 50 \\
9564 & 2001-08-01 06:27 & 14.0 & 21.2 & 130 \\
9567 & 2001-08-02 16:03 & -15.0 & -21.2 & 100 \\
9579 & 2001-08-13 11:11 & -18.0 & -15.1 & 60 \\
9645\tablenotemark{a} & 2001-09-30 12:47 & -19.0 & -28.2 & 210 \\
9651 & 2001-10-03 16:03 & -23.0 & -9.4 & 50 \\
9652 & 2001-10-04 06:23 & 23.0 & -28.9 & 20 \\
9693 & 2001-11-07 08:03 & 11.0 & -17.6 & 60 \\
9725 & 2001-12-01 14:27 & -11.0 & 20.0 & 80 \\
9729 & 2001-12-04 17:36 & 24.0 & -26.5 & 50 \\
9739 & 2001-12-13 11:15 & -13.0 & -10.6 & 520 \\
9746 & 2001-12-19 03:11 & -16.0 & -25.1 & 30 \\
9770 & 2002-01-01 19:15 & 7.0 & 7.0 & 90 \\
9791 & 2002-01-19 17:36 & -4.0 & 3.7 & 140 \\
9812 & 2002-02-02 06:24 & 12.0 & 0.4 & 50 \\
9841 & 2002-02-22 01:36 & -21.0 & 1.9 & 110 \\
9854 & 2002-03-02 09:36 & 10.0 & -17.7 & 30 \\
9858 & 2002-03-03 12:48 & -29.0 & -30.4 & 60 \\
9873 & 2002-03-15 20:48 & -17.0 & -26.9 & 130 \\
9877 & 2002-03-20 03:12 & 18.0 & 11.1 & 260 \\
9894 & 2002-04-03 14:24 & 15.0 & -27.9 & 30 \\
9897 & 2002-04-06 01:36 & -1.0 & -13.9 & 100 \\
9908 & 2002-04-11 19:12 & 5.0 & 3.4 & 70 \\
9924 & 2002-04-23 17:36 & -17.0 & -15.4 & 70 \\
9976 & 2002-05-29 04:48 & -11.0 & -31.2 & 30 \\
9993 & 2002-06-08 17:35 & 7.0 & 2.0 & 50 \\
10006 & 2002-06-17 15:59 & -8.0 & 0.4 & 30 \\
10021 & 2002-07-02 07:59 & -29.0 & -33.2 & 120 \\
10040 & 2002-07-22 01:36 & -23.0 & -19.2 & 20 \\
10049 & 2002-07-25 20:47 & -6.0 & 8.7 & 30 \\
10057 & 2002-07-31 16:03 & -9.0 & -24.7 & 450 \\
10060 & 2002-08-02 17:36 & -29.0 & 2.9 & 50 \\
10078 & 2002-08-13 14:27 & -11.0 & -12.4 & 100 \\
10132\tablenotemark{a} & 2002-09-21 16:03 & 19.0 & -22.1 & 470 \\
10135 & 2002-09-26 11:12 & -27.0 & -32.0 & 60 \\
10152 & 2002-10-10 01:35 & 20.0 & 2.0 & 20 \\
10186 & 2002-11-04 12:48 & 19.0 & -28.0 & 60 \\
10192 & 2002-11-11 03:11 & 13.0 & -26.1 & 630 \\
10193 & 2002-11-12 06:27 & -2.0 & -36.7 & 20 \\
10219 & 2002-12-06 11:12 & -8.0 & -31.7 & 30 \\
10232 & 2002-12-20 14:27 & 13.0 & -17.1 & 20 \\
10253 & 2003-01-10 04:47 & 11.0 & 18.7 & 70 \\
10273 & 2003-01-26 20:47 & 6.0 & 2.4 & 160 \\
10292 & 2003-02-24 19:14 & -8.0 & -8.4 & 150 \\
10298 & 2003-03-03 03:11 & -9.0 & -35.2 & 40 \\
10317 & 2003-03-19 09:39 & 4.0 & -29.7 & 30 \\
10328 & 2003-03-30 08:03 & -11.0 & -33.7 & 10 \\
10327 & 2003-03-31 08:03 & -7.0 & 2.7 & 40 \\
10331 & 2003-04-04 20:48 & -7.0 & -13.0 & 50 \\
10359 & 2003-05-11 20:47 & -16.0 & 3.3 & 20 \\
10417 & 2003-07-19 04:46 & -20.0 & 6.0 & 600 \\
10423 & 2003-07-31 04:47 & -19.0 & -12.1 & 20 \\
10427 & 2003-08-02 19:12 & 3.0 & -7.0 & 110 \\
10428 & 2003-08-04 04:51 & 17.0 & 6.0 & 60 \\
10439 & 2003-08-20 20:47 & 8.0 & 3.0 & 70 \\
10443 & 2003-08-24 06:23 & 22.0 & 4.8 & 20 \\
10453 & 2003-09-03 08:03 & -23.0 & 15.0 & 260 \\
10461 & 2003-09-14 12:48 & 12.0 & 1.3 & 270 \\
10462 & 2003-09-14 22:24 & -9.0 & 3.7 & 90 \\
10480 & 2003-10-14 16:03 & 20.0 & 2.0 & 50 \\
10481 & 2003-10-16 12:47 & -8.0 & -7.9 & 40 \\
10488\tablenotemark{a} & 2003-10-26 11:11 & 8.0 & -32.7 & 930 \\
10492 & 2003-10-26 17:35 & -22.0 & 2.2 & 360 \\
10498\tablenotemark{a} & 2003-11-08 00:03 & -3.0 & 7.0 & 240 \\
10500 & 2003-11-08 20:47 & -8.0 & 1.3 & 40 \\
10503 & 2003-11-15 09:35 & 18.0 & -21.7 & 20 \\
10522 & 2003-12-11 11:11 & 15.0 & 8.9 & 50 \\
10529 & 2003-12-18 22:23 & 9.0 & -17.2 & 20 \\
10532 & 2003-12-23 22:15 & -11.0 & -36.2 & 70 \\
10543 & 2004-01-19 08:03 & -18.0 & -19.6 & 110 \\
10550 & 2004-01-31 20:47 & -8.0 & -21.2 & 40 \\
10553 & 2004-02-03 14:23 & -6.0 & -36.5 & 40 \\
10559\tablenotemark{a} & 2004-02-13 11:15 & 7.0 & 15.7 & 80 \\
10568 & 2004-02-26 20:47 & -17.0 & -13.5 & 30 \\
10591 & 2004-04-11 14:27 & -15.0 & -25.7 & 100 \\
10602 & 2004-04-28 22:23 & -14.0 & 19.7 & 60 \\
10601\tablenotemark{a} & 2004-04-29 14:23 & -9.0 & 5.3 & 310 \\
10605 & 2004-05-03 12:47 & -12.0 & -24.7 & 130 \\
10619 & 2004-05-23 01:36 & -9.0 & 18.5 & 30 \\
10623 & 2004-06-01 00:03 & 7.0 & 3.9 & 70 \\
10626 & 2004-06-04 19:12 & 5.0 & -16.2 & 20 \\
10643 & 2004-07-08 15:59 & -8.0 & -16.4 & 30 \\
10645 & 2004-07-09 03:11 & 12.0 & -26.2 & 20 \\
10671\tablenotemark{a} & 2004-09-06 08:03 & -10.0 & 13.6 & 540 \\
10688 & 2004-10-19 06:27 & -7.0 & 16.6 & 170 \\
10737 & 2005-02-23 03:15 & -9.0 & 12.5 & 50 \\
10753 & 2005-04-12 14:27 & 12.0 & -18.2 & 20 \\
10757 & 2005-04-27 15:59 & -5.0 & -27.2 & 110 \\
10770\tablenotemark{a} & 2005-05-28 23:59 & 12.0 & -17.1 & 70 \\
10771 & 2005-05-29 12:47 & 24.0 & 8.0 & 50 \\
10829 & 2005-12-02 04:48 & 11.0 & -18.7 & 40 \\
10839 & 2005-12-20 11:15 & 18.0 & -30.0 & 40 \\
10846 & 2006-01-14 06:27 & 4.0 & 1.1 & 140 \\
10850 & 2006-01-22 11:12 & 6.0 & -21.6 & 30 \\
10852 & 2006-02-06 22:24 & -10.0 & -28.7 & 20 \\
10868 & 2006-04-04 07:59 & -7.0 & -21.9 & 40 \\
10889 & 2006-05-27 09:39 & -3.0 & -21.4 & 60 \\
10890 & 2006-05-27 22:24 & -14.0 & 7.6 & 50 \\
10902 & 2006-07-30 01:36 & -9.0 & 12.5 & 50 \\
10916 & 2006-10-09 06:24 & -13.0 & -22.2 & 30 \\
10919 & 2006-10-27 03:11 & -16.0 & -17.7 & 20 \\
10937 & 2007-01-07 11:11 & -13.0 & -33.2 & 50 \\
10939\tablenotemark{a} & 2007-01-19 22:27 & -3.0 & 4.2 & 180 \\
10964\tablenotemark{a} & 2007-07-12 09:35 & 4.0 & 0.4 & 90 \\
10971 & 2007-09-27 09:35 & 6.0 & -35.4 & 70 \\
10972\tablenotemark{a} & 2007-10-05 12:51 & -6.0 & -20.7 & 70 \\
10974 & 2007-11-15 20:51 & 12.0 & -15.9 & 40 \\
\enddata 
\tablenotetext{a}{\ Member of the ``Ultra Clean'' subset}
\label{table:PE}  
\end{deluxetable}

\clearpage 
\begin{center}
\begin{deluxetable}{rrrr}  
\tablecolumns{4}
\tablewidth{0pc}
\tablecaption{Identifying Coordinates for No-Emergence Targets.}  
\tablehead{ 
\colhead{Region ID} & \colhead{GONG-Day} & \multicolumn{2}{c}{Ref. Location ($^\circ$)}  \\ 
\colhead{(MDI Orbits)} & \colhead{Reference Date} & \colhead{Lat.} & \colhead{Long.}  
 } 
\startdata 
3137-3140 & 2001-08-05 07:11 & -4.5 & -26.9 \\
3148-3151 & 2001-08-16 08:48 & -13.3 & -11.4 \\
3154-3157 & 2001-08-22 10:23 & -26.5 & -19.4 \\
3175-3178 & 2001-09-12 00:51 & 27.6 & -29.5 \\
3177-3180 & 2001-09-13 19:59 & -15.6 & -26.4 \\
3216-3219 & 2001-10-23 08:47 & 16.0 & -14.1 \\
3225-3228 & 2001-11-01 07:15 & -15.8 & -32.9 \\
3226-3229 & 2001-11-02 00:51 & -20.6 & 12.2 \\
3232-3235 & 2001-11-08 02:27 & -6.7 & 13.6 \\
3234-3237 & 2001-11-10 07:11 & 0.2 & -3.0 \\
3249-3252 & 2001-11-25 00:51 & 0.7 & -23.5 \\
3257-3260 & 2001-12-02 13:39 & 18.5 & 16.7 \\
3277-3280 & 2001-12-23 00:47 & -23.9 & -9.7 \\
3283-3286 & 2001-12-29 10:23 & 25.2 & -21.1 \\
3304-3307 & 2002-01-19 04:00 & -26.7 & -28.0 \\
3355-3358 & 2002-03-10 15:15 & -14.9 & -18.2 \\
3369-3372 & 2002-03-25 04:00 & 18.7 & -9.0 \\
3372-3375 & 2002-03-27 16:48 & -4.4 & -35.0 \\
3405-3408 & 2002-04-29 18:23 & -24.1 & -25.2 \\
3415-3418 & 2002-05-10 00:47 & 5.6 & 11.1 \\
3418-3421 & 2002-05-13 02:22 & 15.6 & -19.2 \\
3430-3433 & 2002-05-25 02:24 & -2.3 & -6.0 \\
3455-3458 & 2002-06-18 16:51 & -8.9 & -17.2 \\
3456-3459 & 2002-06-20 08:48 & -21.1 & -13.0 \\
3471-3474 & 2002-07-05 02:27 & 14.5 & -2.6 \\
3472-3475 & 2002-07-05 13:35 & 12.0 & -23.1 \\
3479-3482 & 2002-07-13 08:47 & -1.4 & 3.0 \\
3484-3487 & 2002-07-18 02:23 & -29.0 & 2.8 \\
3502-3505 & 2002-08-04 18:23 & -21.9 & -10.7 \\
3508-3511 & 2002-08-11 00:47 & -11.4 & -21.4 \\
3519-3522 & 2002-08-22 02:27 & 3.7 & 0.4 \\
3535-3538 & 2002-09-07 05:35 & -8.8 & -36.2 \\
3547-3550 & 2002-09-19 02:23 & 12.9 & -29.0 \\
3555-3558 & 2002-09-27 00:48 & -13.7 & -33.2 \\
3555-3558 & 2002-09-27 02:24 & 15.3 & 7.9 \\
3564-3567 & 2002-10-05 11:59 & -6.0 & -0.1 \\
3565-3568 & 2002-10-07 00:51 & -18.7 & -22.9 \\
3597-3600 & 2002-11-08 07:11 & -29.4 & -32.7 \\
3600-3603 & 2002-11-11 10:24 & -4.5 & -12.8 \\
3607-3610 & 2002-11-17 16:51 & 11.0 & -17.7 \\
3607-3610 & 2002-11-18 07:15 & 6.2 & 7.3 \\
3611-3614 & 2002-11-22 05:36 & 10.4 & 8.5 \\
3615-3618 & 2002-11-26 02:27 & -13.2 & -2.0 \\
3636-3639 & 2002-12-17 04:03 & 2.8 & -35.9 \\
3646-3649 & 2002-12-27 08:48 & -12.5 & -2.7 \\
3683-3686 & 2003-02-02 11:59 & 17.1 & -27.6 \\
3694-3697 & 2003-02-13 05:35 & 11.2 & -6.7 \\
3703-3706 & 2003-02-21 18:26 & -7.7 & 5.9 \\
3703-3706 & 2003-02-21 20:02 & -10.1 & -11.2 \\
3753-3756 & 2003-04-12 18:23 & -6.6 & -8.5 \\
3758-3761 & 2003-04-18 08:47 & -5.9 & -25.4 \\
3780-3783 & 2003-05-10 07:11 & -15.8 & 9.0 \\
3782-3785 & 2003-05-11 18:22 & -13.1 & -27.2 \\
3789-3792 & 2003-05-19 05:35 & -5.4 & -7.5 \\
3795-3798 & 2003-05-24 19:59 & 24.6 & -3.6 \\
3802-3805 & 2003-05-31 11:59 & -20.7 & -33.0 \\
3810-3813 & 2003-06-09 08:47 & -27.6 & 12.6 \\
3855-3858 & 2003-07-24 02:22 & 23.2 & -27.7 \\
3863-3866 & 2003-07-31 13:35 & -7.6 & 17.9 \\
3874-3877 & 2003-08-12 02:27 & 7.9 & 1.0 \\
3893-3896 & 2003-08-30 15:15 & 10.3 & -32.5 \\
3910-3913 & 2003-09-16 21:36 & 15.7 & -4.5 \\
3969-3972 & 2003-11-14 23:11 & 18.4 & 11.1 \\
3983-3986 & 2003-11-28 21:39 & 0.3 & -31.1 \\
3997-4000 & 2003-12-12 15:15 & -12.9 & 1.5 \\
4043-4046 & 2004-01-28 00:47 & -8.4 & 16.4 \\
4044-4047 & 2004-01-29 07:11 & 14.5 & -17.5 \\
4053-4056 & 2004-02-06 20:03 & -18.6 & -23.9 \\
4067-4070 & 2004-02-20 15:12 & -16.7 & 13.1 \\
4073-4076 & 2004-02-26 16:47 & 15.8 & -16.9 \\
4084-4087 & 2004-03-08 21:39 & 1.6 & -25.9 \\
4138-4141 & 2004-05-01 13:35 & -11.8 & -19.7 \\
4157-4160 & 2004-05-20 18:24 & -7.3 & -30.5 \\
4176-4179 & 2004-06-08 13:39 & 21.4 & 11.2 \\
4217-4220 & 2004-07-19 16:47 & 6.4 & -7.0 \\
4228-4231 & 2004-07-30 23:11 & -17.0 & -5.9 \\
4256-4259 & 2004-08-27 15:11 & -10.7 & -16.2 \\
4327-4330 & 2004-11-07 10:23 & -13.6 & -16.2 \\
4334-4337 & 2004-11-13 13:35 & -7.4 & 20.1 \\
4404-4407 & 2005-01-23 00:47 & -6.2 & 2.3 \\
4441-4444 & 2005-02-28 15:15 & 3.6 & -4.9 \\
4518-4521 & 2005-05-17 07:12 & -3.0 & -29.1 \\
4551-4554 & 2005-06-19 00:51 & 10.1 & -24.2 \\
4580-4583 & 2005-07-17 19:59 & 10.8 & 2.5 \\
4604-4607 & 2005-08-10 13:35 & 10.2 & -13.0 \\
4623-4626 & 2005-08-29 15:11 & -8.5 & -25.2 \\
4628-4631 & 2005-09-04 04:03 & 13.3 & -17.5 \\
4669-4672 & 2005-10-15 07:12 & 12.0 & -18.5 \\
4676-4679 & 2005-10-22 08:48 & -5.2 & -27.7 \\
4733-4736 & 2005-12-18 07:12 & 11.8 & -33.0 \\
4833-4836 & 2006-03-28 02:23 & -7.5 & -9.5 \\
4834-4837 & 2006-03-28 16:51 & -11.8 & -25.7 \\
4913-4916 & 2006-06-15 21:35 & -14.2 & -23.1 \\
4916-4919 & 2006-06-19 10:23 & -23.9 & -3.2 \\
4929-4932 & 2006-07-02 05:36 & -10.1 & 19.2 \\
4955-4958 & 2006-07-27 20:03 & -12.7 & 14.9 \\
4959-4962 & 2006-08-01 00:51 & -9.7 & -20.4 \\
5037-5040 & 2006-10-18 05:36 & 9.4 & 6.6 \\
5113-5116 & 2007-01-02 00:51 & -1.6 & 6.5 \\
5233-5236 & 2007-05-01 12:03 & 5.0 & 22.2 \\
5277-5280 & 2007-06-15 05:36 & -2.2 & -8.4 \\
5335-5338 & 2007-08-11 12:03 & 24.7 & -6.2 \\
5374-5377 & 2007-09-19 19:59 & 8.6 & -19.4 \\
5384-5387 & 2007-09-30 02:23 & -10.8 & 5.4 \\
5397-5400 & 2007-10-12 15:11 & -4.0 & -1.4 \\
5411-5414 & 2007-10-26 18:24 & -3.5 & -35.7 \\
5472-5475 & 2007-12-26 18:27 & -12.2 & -34.7 \\

\enddata 
\label{table:NE} 
\end{deluxetable}
\end{center}

\clearpage
\begin{center}
\begin{deluxetable}{cccc}  
\tablecolumns{3}
\tablewidth{0pc}
\tablecaption{Duty Cycle for NE, PE targets} 
\tablehead{ 
\colhead{Time Interval} & \colhead{NE samples} & \colhead{PE samples} & \colhead{Ultra-Clean PE samples} \\
} 
\startdata 
TI-0 & 81 & 89 &  7 \\
TI-1 & 85 & 88 & 10 \\
TI-2 & 85 & 89 & 11 \\
TI-3 & 82 & 87 &  9 \\
TI-4 & 83 & 86 &  9 \\
\enddata 
\label{table:dutycycle} 
\end{deluxetable}
\end{center}

\end{document}